\documentclass[12pt,draftclsnofoot,onecolumn]{IEEEtran}

\usepackage{cite} 
\usepackage{xspace}
\usepackage{filecontents}
\usepackage{float}
\usepackage{pgf, tikz}
\usepackage{graphicx}
\usepackage[cmex10]{amsmath}
\usepackage{mathtools} 

\usepackage{bm}
\usepackage{amssymb} 
\usepackage{array}
\usepackage{amsfonts}
\usepackage{amsthm}   
\usepackage{enumerate} 
\usepackage{enumitem}

\newtheorem{theorem}{Theorem} 

\newtheorem{proposition}{Proposition}
\newtheorem{corollary}{Corollary} 
\newtheorem{remark}{Remark}
\usetikzlibrary{patterns,snakes}

\usepackage{enumerate}
\usepackage{stfloats,epstopdf}
\usepackage{pgfplots}
\usepackage{caption}
\usepackage{subcaption}
\usepackage{tabularx}
\usepackage{pgfplotstable}
\usepgfplotslibrary{dateplot}
\usepackage{cite}
\usepackage{url}
\usepackage{amsbsy}
\usepgfplotslibrary{patchplots}
\usepackage{xcolor}
\usepackage{tikz}
\usetikzlibrary{shadows}
\usetikzlibrary{fadings}
\usetikzlibrary{arrows}
\usetikzlibrary{shapes,snakes}
\pgfplotsset{compat=1.7}
\usetikzlibrary{patterns}
\usetikzlibrary{spy}

\usepackage{amsmath,accents}
\DeclareMathSymbol{\widehatsym}{\mathord}{largesymbols}{"62}
\newcommand\lowerwidehatsym{%
  \text{\smash{\raisebox{-1.3ex}{%
    $\widehatsym$}}}}
\newcommand\fixwidehat[1]{%
  \mathchoice
    {\accentset{\displaystyle\lowerwidehatsym}{#1}}
    {\accentset{\textstyle\lowerwidehatsym}{#1}}
    {\accentset{\scriptstyle\lowerwidehatsym}{#1}}
    {\accentset{\scriptscriptstyle\lowerwidehatsym}{#1}}
}
\DeclareMathSymbol{\widetildesym}{\mathord}{largesymbols}{"65}
\newcommand\lowerwidetildesym{%
  \text{\smash{\raisebox{-1.3ex}{%
    $\widetildesym$}}}}
\newcommand\fixwidetilde[1]{%
  \mathchoice
    {\accentset{\displaystyle\lowerwidetildesym}{#1}}
    {\accentset{\textstyle\lowerwidetildesym}{#1}}
    {\accentset{\scriptstyle\lowerwidetildesym}{#1}}
    {\accentset{\scriptscriptstyle\lowerwidetildesym}{#1}}
}
\pgfplotscreateplotcyclelist{mycyclelist}{
solid, red, line width=1.0pt, every mark/.append style={solid, fill=red}, mark=square, mark size={1}\\%
densely dashed, blue, line width=1.0pt, every mark/.append style={solid, fill=blue}, mark=triangle, mark size={1}\\%
densely dotted, line width=1.0pt, green!70!black, every mark/.append style={solid, fill=green!70!black}, mark=o, mark size={1}\\%
densely dashdotted, line width=1.0pt, brown!70!black, every mark/.append style={solid, fill=brown!70!black}, mark=diamond, mark size={1}\\%
dotted, line width=1.0pt, orange!70!black, every mark/.append style={solid, fill=orange!70!black}, mark=x, mark size={1.3}\\%
thick, solid,  red\\%
densely dashdotted, line width=1.0pt, blue\\%
densely dotted,  line width=1.2pt, green!70!black\\%
solid,  blue, every mark/.append style={solid, fill=blue}, mark=*, mark size={1}\\%
dashed, blue, every mark/.append style={solid, fill=blue}, mark=*, mark size={1}\\%
solid,  red, every mark/.append style={solid, fill=red}, mark=square*, mark size={1}\\%
dashed, red, every mark/.append style={solid, fill=red}, mark=square*, mark size={1}\\%
solid,  green!60!black, every mark/.append style={solid, fill=green!60!black}, mark=triangle*, mark size={1.2}\\%
dashed, green!60!black, every mark/.append style={solid, fill=green!60!black}, mark=triangle*, mark size={1.2}\\%
solid,  brown!70!black, every mark/.append style={solid, fill=brown!70!black}, mark=diamond*, mark size={1.2}\\%
dashed, brown!70!black, every mark/.append style={solid, fill=brown!70!black}, mark=diamond*, mark size={1.2}\\%
}
\pgfplotscreateplotcyclelist{mycyclelist2}{
solid, red, every mark/.append style={solid, fill=red}, mark=square, mark size={1}\\%
densely dashed, blue, line width=1.0pt, every mark/.append style={solid, fill=blue}, mark=triangle, mark size={1}\\%
densely dotted, line width=1.0pt, green!70!black, every mark/.append style={solid, fill=green!70!black}, mark=o, mark size={1.2}\\%
densely dashdotted, line width=1.0pt, brown!70!black, every mark/.append style={solid, fill=brown!70!black}, mark=x, mark size={1.2}\\%
thick, solid,  red\\%
densely dashdotted, line width=1.0pt, blue\\%
thick, solid,  green!70!black\\%
dashed, line width=1.0pt, brown!70!black\\
densely dashdotted, blue, every mark/.append style={solid, fill=blue}, mark=square, mark size={1}\\%
solid, blue, every mark/.append style={solid, fill=blue}, mark=triangle, mark size={1.2}\\%
densely dotted, line width=0.5pt, blue, every mark/.append style={solid, fill=blue}, mark=o, mark size={1}\\%
dashed, blue, every mark/.append style={solid, fill=blue}, mark=x, mark size={1.2}\\%
thick, dashed,  green!70!black\\%
densely dashdotted, red, every mark/.append style={solid, fill=red}, mark=square, mark size={1}\\%
solid, red, every mark/.append style={solid, fill=red}, mark=triangle, mark size={1.2}\\%
densely dotted, red, every mark/.append style={solid, fill=red}, mark=o, mark size={1}\\%
dashed, red, every mark/.append style={solid, fill=red}, mark=x, mark size={1.2}\\%
}
\pgfplotscreateplotcyclelist{mycyclelist3}{
solid,  blue, every mark/.append style={solid, fill=blue}, mark=*, mark size={1}\\%
densely dashed, red, every mark/.append style={solid, fill=red}, mark=x, mark size={1.5}\\%
}

\newcommand{\AuthorOne}{Sheeraz A. Alvi, \textit{Member, IEEE}} 
\newcommand{\AuthorTwo}{Xiangyun Zhou, \textit{Senior Member, IEEE}}
\newcommand{\AuthorThree}{Salman Durrani, \textit{Senior Member,~IEEE}}
\newcommand{\ThankOne}{Sheeraz Alvi, Xiangyun Zhou, Salman Durrani are with the Research School of Engineering, the
	Australian National University, Canberra, ACT 2601, Australia
	(emails: \{sheeraz.alvi, xiangyun.zhou, salman.durrani\}@anu.edu.au).}
\newcommand{\ThankTwo}{This paper is presented in part at IEEE Globecom 2018 \cite{sheeraz-2018}.}

\begin{document}
%
\title{Optimal Compression and Transmission Rate Control for Node-Lifetime Maximization}
\author{\IEEEauthorblockN{\AuthorOne,~\AuthorTwo, and~\AuthorThree\thanks{\ThankOne}\thanks{\ThankTwo}}}
\maketitle


\vspace{-1.5cm}
\begin{abstract}
\normalsize


We consider a system that is composed of an energy constrained sensor node and a sink node, and devise optimal data compression and transmission policies with an objective to prolong the lifetime of the sensor node.
%
%
While applying compression before transmission reduces the energy consumption of transmitting the sensed data, blindly applying too much compression may even exceed the cost of transmitting raw data, thereby losing its purpose. Hence, it is important to investigate the trade-off between data compression and transmission energy costs.
%
%
In this paper, we study the joint optimal compression-transmission design in three scenarios which differ in terms of the available channel information at the sensor node, and cover a wide range of practical situations. We formulate and solve joint optimization problems aiming to maximize the lifetime of the sensor node whilst satisfying specific delay and bit error rate (BER) constraints.
%
%
Our results show that a jointly optimized compression-transmission policy achieves significantly longer lifetime ($90\%$ to $2000\%$) as compared to optimizing transmission only without compression. Importantly, this performance advantage is most profound when the delay constraint is stringent, which demonstrates its suitability for low latency communication in future wireless networks.
\end{abstract}

\IEEEpeerreviewmaketitle

\vspace{-0.25cm}
\begin{IEEEkeywords}
\normalsize \vspace{-0.25cm}
Machine-type communication, lifetime, energy efficiency, data compression, data transmission.
\end{IEEEkeywords}
{\enlargethispage*{0.5 cm}}

\newpage

\section{Introduction}

The notion of Internet of things (IoT) calls for novel solutions to realize wireless connectivity across heterogeneous and autonomous wireless devices such as sensors, actuators, etc., which are often referred to as machine-type communication (MTC) devices \cite{fuqaha-2015,dawy-2017}. The sensor based MTC devices within an IoT system, are supposed to acquire physical information from the environment and transmit it to a central data fusion station while satisfying stringent technical requirements in terms of maximal energy efficiency, ultra-low latency, and application specific rigorous data reliability.

Due to their wireless and unattended operation, MTC devices are mostly battery operated and are severely energy constrained. Thus, prolonging the lifetime of these sensor based MTC devices, which is defined as the time taken by the MTC device to deplete all of its energy, is of paramount importance \cite{hanzo-2017}. In the existing literature, the lifetime maximization problem has been approached from different perspectives such as green channel access, sleep-wake scheduling, coverage, efficient routing, network coding, data aggregation, see \cite{hanzo-2017} and the references therein.

The low-cost and miniature sized CMOS cameras and microphones have made it possible to acquire multimedia information, i.e., image, audio, and video, from the environment enabling the notion of wireless multimedia sensor networks (WMSN) \cite{ian2007} and Internet of multimedia things (IoMT) \cite{iomt-2015}. In most of the applications of WMSN and IoMT, the amount of sensed data (raw data) can sometimes be very large, resulting in high transmission cost. In this regard, data compression schemes have been proposed \cite{compressionSurvey, compression1, compression2, compression3}, which decrease the amount of data to be transmitted and thus alleviate the transmission energy cost. Typically, the energy cost of compression and transmission is around 15\% and 80\% of the total energy consumed by the sensor node, respectively \cite{Jung-2009, raghunathan2002energy}. The transmission cost depends upon the required transmission rate and the signal strength. Unlike the transmission energy cost which linearly increases with the size of data to be transmitted, the compression energy cost has a non-linear relationship with the compression ratio \cite{tahir2013cross}. Owing to this non-linearity, blindly applying too much compression may even exceed the cost of transmitting raw data, thereby losing its purpose \cite{sadler2006data}.

From the compressive sensing perspective, \cite{R1} considered a wireless powered cognitive radio network and optimized the time slot allocation for energy harvesting, sensing, and transmission processes. By using compressive sensing, less number of samples are collected which reduces the sensing cost as well as the data transmission cost. The reader is referred to \cite{R2} for an overview of related works on compression and transmission frame design using compressive sensing in wireless sensor networks.

\textbf{Design Challenge}: Existing adaptive transmission rate control schemes \cite{rate, UYSAL-2004, Zafer-2008} attempt to prolong node-lifetime without considering data compression \cite{compressionSurvey, compression1, compression2, compression3}. In this work, our focus is to devise joint data compression-transmission policies to optimally utilize the energy resources to maximize the lifetime of an energy constrained sensor based MTC device.

Prior works on power-rate adaptation \cite{UYSAL-2002, UYSAL-2004, Zafer-2008, Berry-2002, Zafer-2009}, which do not employ data compression, have considered transmission energy as a monotonically increasing function of the transmission rate. Therefore, these schemes propose to transmit data at lower transmission rates under given delay constraint to achieve energy efficiency. These schemes assume the distance between communicating devices is large, thus the transmit power dominates the circuit power \cite{cui2005energy, geoffrey-2011}. However, in many practical sensor networks, e.g., body area networks, the distance is fairly small and the circuit power cost cannot be ignored. In this regard, considering the power amplifier cost as a function of the constellation size and transmit power, the transmission energy is not anymore a monotonically increasing function of the transmission rate, as shown in \cite{cui2005energy}. This is the case particularly for smaller constellation sizes, which are more common in sensor networks. Therefore, simply decreasing the transmission rate may not necessarily improve energy efficiency.

Generally, MTC poses application specific quality of service (QoS) requirements in terms of data reliability. Therefore, in this work we devise compression and transmission policies  whilst guaranteeing the specific delay constraint and bit error rate (BER) requirement, to ensure data freshness and data integrity. In addition, we adopt radio duty cycling (RDC) which is a well known mechanism to achieve energy efficiency \cite{Jung-2009}, it has been considered separately in the literature for network lifetime maximization. Moreover, we also adopt the recent concept of deep sleep which allows MTC devices to switch their micro-controller unit (MCU) from active to an inactive state in which minimal processing resources are available to allow data sensing, while the radio is kept inactive too.

\subsection{Paper contributions}

We consider a monitoring system, in which an energy constrained sensor node acquires some physical information from its vicinity, applies data compression on the sensed data, and transmits the compressed data to the sink node. The sink node may be able to feedback perfect or imperfect channel information to the sensor node, depending upon the considered scenario. Accordingly, the sensor node devises an optimal compression and transmission policy based on the available channel information with an objective to maximize its lifetime. To the best of our knowledge, this is the first work in which the data compression and transmission are jointly designed and optimized. In contrast, prior studies only considered the optimization of transmission policy without compression. As shown in this work, the joint optimization of compression and transmission results in a substantial improvement in the lifetime of the sensor node. To this end, we consider three scenarios which differ in terms of the available channel information at the sensor node. The first scenario assumes the availability of the perfect instantaneous channel gain information at the sensor node and provides the benchmark theoretical performance. The remaining two scenarios assume the availability of quantized channel gain and statistical channel gain information,\footnote{In \cite{sheeraz-2018}, we consider only statistical channel gain information available at the sensor node.} respectively, at the sensor node and provide the performance of practical wireless sensor based MTC systems.

Our investigation leads to the following observations and design guidelines:

\begin{itemize}

\item Our results show that a jointly optimized compression-transmission policy performs much better than optimizing transmission only without compression under any given BER and delay constraints. The performance gain observed ranges from $90\%$ to $2000\%$ and is most profound when the delay constraint is stringent.

\item The optimal level of compression is insensitive to the change in the BER requirement. However, the optimal transmission rate increases as BER constraint gets less stringent.

\item The best strategy is to reduce compression and increase the transmission rate when the delay constraint gets more stringent and vice versa. The optimal level of compression and transmission rate are more sensitive to the delay constraint when the system requires low latency, while they remain roughly unchanged when the delay constraint is relaxed beyond a certain point.

\end{itemize}

\textbf{Notations}: $\mathbb{P}\{\cdot\}$ and $\mathbb{E}\{\cdot\}$ represent the probability and expectation operators, respectively. $\exp(\cdot)$ represents the exponential. $\lfloor{\cdot}\rfloor$ and $\left \lceil{\cdot} \right \rceil$ is the floor and ceil operations, respectively. $|\cdot|$ represents the absolute value. $\nabla\{\cdot\}$ is the gradient operator. $[\cdot]^{\top}$ is the transpose operator.

The rest of the paper is organized as follows. The system model is presented in Section~II. The communication scenarios, which differ in the level of the channel knowledge at the
sensor node, are discussed in Section~III. The lifetime maximization problems and their solutions are provided in Section~IV. Results are presented in Section~V. Finally, Section~VI
concludes the paper.

\begin{figure}[t]
  \centering
  \includegraphics[scale=0.90]{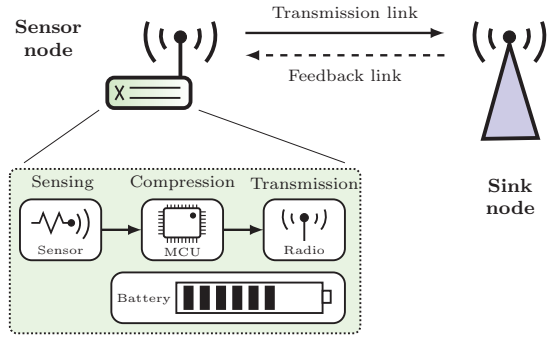}\\
  \caption{Illustration of the considered system model, comprising a sensor and a sink node.}
  \label{fig-sys}
  \vspace{-0.3cm}
\end{figure}

\section{System model}\label{sec-system}

\begin{figure}[t]
  \centering
  \includegraphics[scale=0.87]{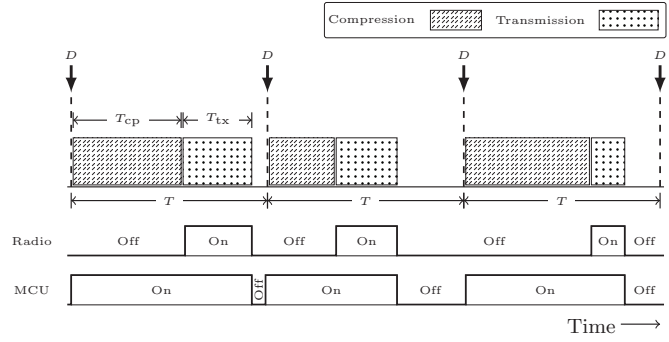}\\
  \caption{Timing diagram for compression and transmission processes and corresponding radio and MCU activity cycles.}
  \label{fig-timing}
  \vspace{-0.3cm}
\end{figure}

We consider a system consisting of a sensor based MTC device (sensor node) which is periodically transmitting its sensed data to a sink node, as illustrated in Fig. \ref{fig-sys}. Both nodes are equipped with a single omnidirectional antenna. The sensor node is battery operated and energy constrained, whereas the sink node has no energy constraint. The system follows a block-wise operation with a block of duration $T$, as shown in Fig. \ref{fig-timing}. Within each time block the sensor node performs three main functions, i.e., (i) sensing, (ii) compression, and (iii) transmission, each having individual completion time and energy cost. The block-wise operation and the duration of the block length depends on the sensor application. For example, in WMSN and IoMT applications, the fresh data is periodically available which needs to be transmitted within a given deadline.

For energy efficient operation, we employ RDC, i.e., radio is kept in the inactive state except during the transmission process. Moreover, MCU is kept in the inactive state, when it is neither compressing nor transmitting data, referred to as deep sleep. The transition periods from active to inactive states and vice versa are fast enough to be negligible for both radio and MCU. We assume the power consumed by the radio and MCU in inactive states is negligible  \cite{raghunathan2002energy, Jung-2009}.

\textbf{Sensing:}
The sensing operation is as follows. Firstly, the sensor node acquires the required information from the physical environment and encodes it to a data of size $D$ bits. The sensed data during a given time block, is available for transmission at the start of next time block. In this work, we make the following assumptions regarding data sensing:
\begin{itemize}
  \item The data sensing can be done in parallel while the compression and transmission processes are being executed.
  \item The sensed data is always periodically available, which is in line with prior works in this area \cite{UYSAL-2002,UYSAL-2011,Zafer-2009,Gregori-2013,wanchun-2016,salman-book}.
  \item The periodic sensing to acquire a fixed amount of physical information from the environment typically consumes a constant time and energy \cite{mao-2014}.
  \item The amount of data to be sensed, and the associated cost of sensing, depend only on the application and is independent of the compression and transmission processes.
\end{itemize}

Let the time and power spent by the sensor node to sense data of size $D$ bits be denoted by $T_{\textup{sen}}$ and $P_{\textup{sen}}$, respectively.

\textbf{Compression:}
Before transmission, the sensed data of size $D$ bits is compressed into $D_{\textup{cp}}$ bits as per the given compression ratio $\frac{D_{\textup{cp}}}{D}$. The \emph{compression time}, denoted by $T_{\textup{cp}}$, is defined as the time required to compress raw data, $D$, into compressed data, $D_{\textup{cp}}$. A non-linear compression cost model is proposed in \cite{tahir2013cross} to evaluate the compression cost. This model is validated for JPEG and JPEG2000 compression algorithms in \cite{tahir2013cross}. In this work, we adopt this non-linear model to compute the compression time as a function of the compression ratio, $\frac{D_{\textup{cp}}}{D}$, as
\begin{equation}\label{t-comp}
  T_{\textup{cp}} =  \tau D \bigg( \Big(\frac{D}{D_{\textup{cp}}} \Big)^\beta - 1 \bigg),
\end{equation}
\noindent where $\beta$ is the compression algorithm dependent parameter and $\tau$ is the per bit processing time. In general, $\beta$ is proportional to the compression algorithm complexity and it determines the time cost for achieving a given compression ratio for given hardware resources. $\beta$ can be calculated off-line for any specified compression algorithm and given hardware resources. $\tau$ depends upon the MCU processing resources and the number of program instructions executed to process 1 bit of data. Note that $\tau$ does not represent the compression time per bit. It can be given as
\begin{equation}\label{t-b-c}
 \tau =  \underbrace{\frac{ \textrm{instructions}}{\textrm{program}}}_{(i)} \times \underbrace{\frac{\textrm{clocks}}{\textrm{instruction}}}_{(ii)} \times \underbrace{\frac{\textrm{seconds}}{\textrm{clock}}}_{(iii)} \times \underbrace{\frac{1}{\textrm{reg}}}_{(iv)},
\end{equation}
\noindent The explanation for the terms in \eqref{t-b-c} is as follows:
\begin{description}
  \item [$(i)$] We assume a single-instruction program that is able to process 1 bit of information.
  \item [$(ii)$] Most instructions in a typical sensor mote MCU are executed in 1 clock cycle. We assume a single instruction is executed in 1 clock cycle.
  \item [$(iii)$] Seconds per clock represents the clock speed, i.e., the inverse of the MCU operational frequency which typically is between few MHz to hundreds of MHz.
  \item [$(iv)$] $\textrm{reg}$ represents MCU register size and its value for typical sensor motes is 8-bit. For 8-bit processor, the execution time to process 1 bit or up to 8 bits is the same. We assume $D$ is large (thousands of bits) and it will be processed in chunks of 8 bits.
\end{description}

Let $P_{\textup{cp}}$ denote the power consumed by the sensor node during data compression process. $P_{\textup{cp}}$ is the same as the power consumed while sensor mote's MCU is processing information. Its value is predefined for a given sensor mote with a given hardware processing capability.

\textbf{Transmission:}
Once the compression process is complete, the sensor node needs to transmit the compressed data, $D_{\textup{cp}}$, within the next $T-T_{\textup{cp}}$ seconds. The sensed data needs to be compressed and transmitted within each time block, hence the delay constraint is $T$ seconds. The \emph{transmission time}, denoted by $T_{\textup{tx}}$, depends upon the compressed data size, $D_{\textup{cp}}$, and the link transmission rate, $r$. We consider the sensor node uses $M$-QAM modulation scheme with constellation size equal to $M=2^l$, where $l={1,2,3,..,L}$. Thereby, $T_{\textup{tx}}$ is given as
\begin{equation}\label{t-comm}
  T_{\textup{tx}} =  \frac{D_{\textup{cp}}}{r},
\end{equation}
\noindent where
\begin{equation}\label{r}
r =  \frac{\log_2 \big(  {M} \big) }{T_\textup{s}},
\end{equation}
\noindent where $T_\textup{s}$ is the symbol period and $M$ is the modulation constellation size.

To compute the data transmission power cost, denoted by $P_\textup{tx}$, we adopt a practical model as given in \cite{cui2005energy}. In this model, the total power cost of the sensor node while transmitting data, denoted by $P_\textup{tx}$, is divided into three main components, the transmitted power, RF power amplifier, and the communication module circuitry power, denoted by $P_\textup{t}$, $P_\textup{amp}$, and $P_\textup{o}$, respectively. Accordingly, $P_\textup{tx}$ is given as follows
\begin{equation}\label{p-tx}
 P_\textup{tx} =  P_\textup{t} + P_\textup{amp} + P_\textup{o}.
\end{equation}
$P_\textup{o}$ is further divided into different communication circuitry modules \cite{cui2005energy}
\begin{equation}\label{p_o}
 P_\textup{o} =  P_\textup{fil} + P_\textup{mix} + P_\textup{syn},
\end{equation}
\noindent where $P_\textup{fil}, P_\textup{mix},$ and $P_\textup{syn}$ is the power consumed by filter, mixer, and frequency synthesizer, respectively. The $P_\textup{amp}$ is a function of transmitted power and it can be given as \cite{lee2003design}
\begin{equation}\label{p-amp}
 P_\textup{amp} =  \Big(\frac{\varepsilon}{\mu}-1 \Big) P_\textup{t},
\end{equation}
\noindent where $\mu$ represents the drain efficiency of the power amplifier and $\varepsilon$ represents the peak-to-average ratio (PAR) which depends upon the modulation scheme and the associated constellation size. Since, we consider $M$-QAM modulation, $\varepsilon$ is given as \cite{meyr1998digital}
\begin{equation}\label{mqam-par}
 \varepsilon =  3 \frac{{M}^{\frac{1}{2}}-1}{{M}^{\frac{1}{2}}+1}.
\end{equation}
Substituting the value of $P_\textup{amp}$ in \eqref{p-tx} and re-arranging, $P_\textup{tx}$ can be rewritten as
\begin{equation}\label{p-comm}
 P_\textup{tx} =  \frac{\varepsilon}{\mu} P_\textup{t} + P_\textup{o}.
\end{equation}

We assume that the battery used for sensor node possesses a limited charge storage capacity as well as a maximum current withdrawal limit. Therefore, instantaneous power demand of any process at any state should not exceed the maximum allowable limit. Specifically, the transmission power cost, which is a function of transmit power, needs to meet this power bound in order to ensure the feasibility of the system for practical sensor networks.

\textbf{Channel model:}
The sensor node is located at a distance $d$ from the sink node. The channel between the two nodes is composed of a large-scale path loss, with path loss exponent $\alpha$, and small-scale quasi-static flat Rayleigh fading channel, i.e., the fading channel coefficient $h$ remains constant over a time block and is independently and identically distributed from one time block to the next. The additive noise is assumed to be AWGN with zero mean and variance $\sigma^2$.

The $pdf$ of the instantaneous channel gain, $|h|^2$, is exponentially distributed and is given as
\begin{equation}\label{avg-avg-p}
f\big(|h|^2\big) \triangleq \frac{1}{\varsigma} \exp \bigg( -\frac{|h|^2}{\varsigma} \bigg), \quad |h|^2 \geqslant 0,
\end{equation}
\noindent where $\varsigma$ represents the scale parameter of the probability distribution.

\textbf{Node-Lifetime:}
We assume that the sensor node's battery is initially fully charged. Based on the battery capacity, operating voltage, and rate of energy consumption, we can calculate the node-lifetime, denoted by $T_{\textup{NL}}$, which is defined as the time taken by the node to deplete all of its battery energy. The node-lifetime, $T_{\textup{NL}}$, can be given as
\begin{equation}\label{t-life}
  T_{\textup{NL}} =  \frac{B_{\textup{cap}}V_{\textup{op}}}{  P_{\textup{avg}}},
\end{equation}
\noindent where $B_{\textup{cap}}$ represents the battery capacity that is a measure of the charge stored by the battery, $V_{\textup{op}}$ is the operating voltage, and $P_{\textup{avg}}$ represents the average power consumption by the sensing, compression, and transmission processes and is given as
\begin{equation}\label{p-avg}
  P_{\textup{avg}} =  \frac{  T_{\textup{sen}}P_{\textup{sen}} + \mathbb{E} \big[ \Psi \big ] } {T},
\end{equation}
\noindent where $\mathbb{E}[\cdot]$ is the expectation operator and $\Psi$ is the energy consumed by the compression and transmission processes in a given time block which is given as
\begin{equation}\label{psi}
\Psi~=~T_{\textup{cp}}P_{\textup{cp}} + T_{\textup{tx}} P_{\textup{tx}}.
\end{equation}
Note that the compression and transmission energy costs may change from one time block to the next. However, the sensing energy cost, $T_{\textup{sen}}P_{\textup{sen}}$, is the same for each time block.

\section{Communication Scenarios and Channel Knowledge}

The level of channel information available at the sensor node changes the compression and transmission policy design, since it imposes different constraints on the system which needs to comply with the channel knowledge. We consider three scenarios serving different important purposes. The perfect instantaneous channel gain information (CGI) (Scenario~1) is commonly used for theoretical performance analysis as a benchmark in wireless sensor networks. In reality when a feedback link with limited bandwidth is available then a quantized (imperfect) CGI knowledge is shared (Scenario~2). Finally, when the feedback link is not available, then the sensor node relies on the statistical channel information (Scenario~3).
The considered scenarios are summarized as follows:

\begin{enumerate}[leftmargin=*,labelindent=16pt,label= Scenario \arabic*:]
  \item Perfect CGI is available at the sensor node, when perfect feedback is available.
  \item Imperfect CGI is available at the sensor node, when limited feedback is available.
  \item Statistical CGI is available at the sensor node, when there is no feedback.
\end{enumerate}

Note that the instantaneous CGI is assumed to be known at the sink node, which is a reasonable assumption when the sink node has no constraint on energy and data processing capability \cite{R13}.

\subsection{Instantaneous CGI available at the sensor node}\label{sec-CGI}

The sensor node is able to adaptively control the compression and transmission rate if the instantaneous CGI is available. Therein, we consider the following two scenarios.

\emph{Scenario 1}: In this scenario, perfect instantaneous CGI is available at the sensor node. Therein, at the start of each time block, the sink node perfectly estimates the instantaneous fading channel coefficient and computes the CGI, $|h|^2$. The sink node then feeds back this CGI to the sensor node perfectly. Hence, perfect knowledge of instantaneous CGI is available at the sensor node.

\emph{Scenario 2}: In this scenario, imperfect instantaneous CGI is available at the sensor node. Therein, at the start of each time block, the sink node perfectly estimates the instantaneous fading channel coefficient and computes the CGI, $|h|^2$. Moreover, the sink node is required to quantize the CGI to bound the feedback overhead, since in practical systems only limited feedback is available. Note, the CGI is a real and positive value which allows efficient quantization using a small number of bits \cite{yoo2007multi}. The quantization of the CGI at the sink node, due to limited feedback, results in channel uncertainty at the sensor node side. We assume a perfect feedback link, hence CGI available at the sensor node is only subject to imperfection due to the quantization process.

The actual instantaneous CGI, $|h|^2$, can fall anywhere in the range $[0,\infty)$. We divide this range into $2^B$ quantization intervals, where $B$ represents the number of feedback bits. The range of each of these quantization intervals is selected such that the probability of instantaneous CGI, $|h|^2$, falling in any of the given intervals is the same. Corresponding to these intervals, let the set of quantization levels be denoted by $\mathcal{C}=\{c_1, c_2,...,c_{2^{B}}, c_{2^{B}+1}\}$, where $c_j$ represents the $j$th quantization level and $c_1=0$ and $c_{2^{B}+1}=\infty$. For a given CGI, $|h|^2$, the sink node determines the interval $[c_i,c_{i+1})$, for $i \in \{1,2,...,2^{B}\}$ such that $c_i \leqslant |h|^2 < c_{i+1}$, and feeds back the index $i$ corresponding to the quantization interval $[c_i,c_{i+1})$ to the sensor node. Note that the CGI, $|h|^2$, falls in each of these intervals with equal probability, i.e.,
\begin{equation}
\mathbb{P} \big\{ c_i \leqslant |h|^2 < c_{i+1} \big\} = \frac {1} {2^{B}}, \quad \forall \; i \in \{1,2,...,2^{B}\},
\end{equation}

\noindent where $\mathbb{P}\{\cdot\}$ represents the probability. We assume the time spent in estimation, quantization, and feedback is negligible, thus the sensor node can effectively exploit the provided CGI.

\subsection{Statistical CGI available at the sensor node}

\emph{Scenario 3}: In this scenario, the sink node has perfect estimate of the channel but no instantaneous feedback is available. Only the statistical CGI is available at the sensor node. Therefore, the sensor node cannot adapt compression and transmission policies to varying channel conditions in different time blocks. Instead a constant compression ratio and transmission rate is determined, and subsequently used in each time block, which maximizes the node-lifetime.

\subsection{BER Expression}

We consider the sensor node is able to control the transmission rate, wherein the best $M$ value needs to be determined which will allow the sensor node to achieve the required system performance. Note that many different BER expressions exist in the literature for $M$-QAM. Here, we use the following BER bound defined for $M$-QAM modulation scheme \cite{foschini1983digital}, since it is easy to invert in order to obtain $M$ as a function of the required BER
\begin{equation}\label{ber}
  \textup{BER} \leqslant \omega_2 \; \textup{exp} \bigg( -  \frac { \omega_1} {(M-1)} \gamma \bigg),
\end{equation}

\noindent where $\omega_1$, $\omega_2$ are constants and $\gamma$ represents the received signal-to-noise ratio (SNR) which is defined as follows \cite{goldsmith2005wireless}
\begin{equation}\label{snr}
 \gamma = \kappa \frac {P_\textup{t}|h|^2}{\sigma^2 d^\alpha},
\end{equation}

\noindent where $\kappa=\big(\frac{\lambda}{4\pi}\big)^2$ is the attenuation factor, $\lambda$ is the wavelength and $P_\textup{t}$ is the transmit power. For $M \geqslant 4$ and $0 \leqslant \gamma \leqslant 20\textup{ dB}$, the bound in \eqref{ber} with $\omega_1=1.5$ and $\omega_2=0.2$, is tight to within $1$~dB of the exact result in \cite{goldsmith2005wireless}.

In Section \ref{sec-node-life}, we will use \eqref{ber} to determine the optimal design parameters for each of the scenarios defined above.

\section{Node-Lifetime Maximization Problem}\label{sec-node-life}

The main problem we address is to determine the optimal compression and transmission policies which will maximize the node-lifetime under given delay constraint and BER performance. From \eqref{t-life}, we can see that the node-lifetime is calculated using a predefined initial energy level, $B_{\textup{cap}} V_{\textup{op}}$, and the controllable rate of energy consumption, $P_{\textup{avg}}$, which is a function of $\mathbb{E} [\Psi]$. Node-lifetime is inversely proportional to $\mathbb{E} [\Psi]$, which implies that maximizing node-lifetime is equivalent to minimizing $\mathbb{E} [\Psi]$. Based on the compression and transmission energy cost models, defined in Section~II, $\Psi$ inherits the tradeoff between data compression and transmission.

In this section, we study three different problems based on the scenarios defined in Section~\ref{sec-CGI}. In Scenario~1 and Scenario~2, the knowledge of the instantaneous CGI is used by the sensor node to optimally choose the design parameters in order to adapt to each realization of the channel, with an objective to minimize $\Psi$ in each time block. On the other hand, in Scenario~3, when no knowledge of the instantaneous CGI is available, the optimal design parameters are set to be fixed for each time block, wherein, the objective is still to minimize $\Psi$. It is because, in Scenario~3, the value of $\Psi$ is the same for all time blocks.

\subsection{Instantaneous CGI available at the sensor node}

In this subsection, we consider an adaptive compression and transmission rate control system in which for a given time block, the design target is to minimize the energy cost of compression and transmission under the given delay and BER constraints. Based on the availability of the instantaneous CGI at the sensor node we study the following two problems.

\subsubsection{Perfect CGI availability}

The first problem we study, considering the availability of perfect CGI at the sensor node, can be summarized as follows:

Problem~1: What is the optimal compression and transmission policy that minimizes the compression and transmission energy cost under specific delay and BER constraints, when perfect channel gain is known at the sensor node?

The energy consumption cost of compression and transmission, $\Psi$, is defined in \eqref{psi}. Now given the value of instantaneous CGI, $|h|^2$, Problem~1 can be expressed as follows
\begin{equation}\label{opt-prob-p-1}
\begin{aligned}
& \underset{ M, P_{\textup{t}}, D_{\textup{cp}}} {\textup{minimize}}
& & \Psi (M, P_{\textup{t}}, D_{\textup{cp}})                           \\
& \textup{subject to}
& &   T_{\textup{cp}} +  T_{\textup{tx}} \leqslant T,  \quad \textup{BER}(M, P_{\textup{t}} ) \leqslant \phi,     \\
& & & P_{\textup{t}} \geqslant 0, \quad M \geqslant 2,  \quad M \leqslant M_\textup{max},   \\
& & & D_{\textup{cp}} \geqslant D_{\textup{min}}, \quad D_{\textup{cp}} \leqslant D.
\end{aligned}
\end{equation}

\noindent where the first constraint defines the delay constraint for the data delivery, thus both compression and transmission processes should be completed within the deadline and the second constraint mandates that the $\textup{BER}$ should be below or equal to a specific threshold value denoted by $\phi$. The remaining constraints reflect practical range of values for $ M, P_{\textup{t}}$, and $D_{\textup{cp}}$. In the fourth constraint, $M_\textup{max}$ is the maximum value of the constellation size which can be used by the sensor node. In the fifth constraint, $D_{\textup{min}}$ represents the maximum achievable compression that can be applied using a given compression algorithm and the compressed data is completely transmitted within the same time block at the highest allowed transmission rate. Thus, $D_{\textup{min}}$ is the lower bound on $D_{\textup{cp}}$ and a function of $M_\textup{max}$, $T$, $D$ and $\beta$. $D_{\textup{min}}$ is computed by numerically solving the following equation using the variable-precision floating-point arithmetic (VPA) method
\begin{equation}\label{Dmin}
\tau D^{\beta+1} D_{\textup{min}}^{-\beta} - \tau D  +  \frac{ D_{\textup{min}} T_\textup{s} \ln(2) }{  \ln(M_\textup{max}) } = T.
\end{equation}

In order to solve \eqref{opt-prob-p-1}, we first present Proposition~1, which allows the solution to \eqref{opt-prob-p-1} to be given by Theorem 1.

\begin{proposition}\label{propo-1}
The optimal $P_{\textup{t}}$ to minimize $\Psi(M, P_{\textup{t}}, D_{\textup{cp}})$ for given values of $M$ and $D_{\textup{cp}}$ while satisfying the constraints in \eqref{opt-prob-p-1} is given by
\begin{equation}\label{min-Pt-propo-1}
   P_\textup{t} = (1 - M) \frac { \Omega } {|h|^2},
\end{equation}
\noindent where
\begin{equation}
\Omega = \frac { \sigma^2 d^\alpha \ln( \phi/\omega_2 )} {\omega_1 \kappa}.
\end{equation}
\end{proposition}
\begin{IEEEproof}
The proof is provided in Appendix~\ref{A}.
\end{IEEEproof}

Using the result in Proposition 1, substituting $T_{\textup{cp}}$, $T_{\textup{tx}}$, $r$, $P_\textup{tx}$, $\varepsilon$ and $P_{\textup{t}}$ from \eqref{t-comp}, \eqref{t-comm}, \eqref{r}, \eqref{p-comm}, \eqref{mqam-par} and  \eqref{min-Pt-propo-1}, respectively, in \eqref{psi} yields $\Psi$ as a function of $M$ and $D_{\textup{cp}}$ for given instantaneous CGI, $|h|^2$, which can be expressed as follows
\begin{equation}\label{psi-dc-1}
  \Psi(M,D_{\textup{cp}})  =    \tau D^{\beta+1} D_{\textup{cp}}^{-\beta} P_{\textup{cp}} - \tau D P_{\textup{cp}}
                                -  \frac{ D_{\textup{cp}} T_\textup{s} \ln(2) }{  \ln(M) }  \Bigg(   \frac {3 \Omega ({M}^{\frac{1}{2}} - 1)^2 }  {\mu  |h|^2} - P_\textup{o} \Bigg).
\end{equation}

Now a simpler equivalent optimization problem with only two design parameters, i.e., $M, D_{\textup{cp}}$, needs to be solved and the third parameter $P_\textup{t}$ can be obtained using the result in Proposition 1. Accordingly, the solution to the optimization problem in \eqref{opt-prob-p-1} is given by the following theorem.

\begin{theorem}

In solving the optimization problem in \eqref{opt-prob-p-1}, the optimal constellation size is given by the following conditional expression
\begin{equation}\label{optimal-M-1}
M^*=
    \begin{cases}
      \fixwidetilde{M},    &   \textup{if} \; \mathcal{Q}(\fixwidetilde{M},\fixwidetilde{D}_\textup{cp}) < T. \\
      \min  \big( \fixwidehat{M}, M_\textup{max} \big),      &   \textup{otherwise.}
    \end{cases}
\end{equation}
\noindent where $\fixwidetilde{M}$ and $\fixwidehat{M}$ are given by the solution of the following equations which can be solved numerically using the \textup{VPA} method
\begin{equation}\label{optimal-M-tilde}
 \frac { 3 \Omega  } {\mu |h|^2}
  \big( \fixwidetilde{M}^{\frac{1}{2}} - 1 \big) \big( (\ln(\fixwidetilde{M})-1) \fixwidetilde{M}^{\frac{1}{2}} + 1 \big)  + P_\textup{o}  =  0,
\end{equation}
\begin{equation}\label{optimal-M-bar}
\frac{T}{D} + \tau - \tau
                       \xi^{\frac{-\beta}{\beta+1}}
  =  \frac{ T_{\textup{s}} \ln(2) } {\ln\big(\fixwidehat{M}\big)}
                      \xi^{\frac{1}{\beta+1}},
\end{equation}
\noindent where
\begin{equation}\label{xi}
\xi =          \frac {
                             P_\textup{cp}  -   P_\textup{o}
                             - \frac { 3 \Omega  } {\mu |h|^2}  \big( \fixwidehat{M}^{\frac{1}{2}} {-} 1 \big) \big( (\ln(\fixwidehat{M}) {-} 1) \fixwidehat{M}^{\frac{1}{2}} + 1 \big)
                      }
                      {
                        \frac { 3 \Omega  } {\tau \beta \mu |h|^2}
                        \big( \fixwidehat{M}^{\frac{1}{2}} - \fixwidehat{M} \big)
                      },
\end{equation}
\noindent respectively,
\begin{equation}\label{optimal-Q-1}
\mathcal{Q}(\fixwidetilde{M},\fixwidetilde{D}_\textup{cp}) \triangleq \tau D^{\beta+1} \fixwidetilde{D}_{\textup{cp}}^{-\beta} - \tau D +  \frac{ \fixwidetilde{D}_{\textup{cp}} T_\textup{s} }{  \log_2 \big( \fixwidetilde{M} \big) },
\end{equation}
\noindent and
\begin{equation}\label{optimal-Dcp-tilde}
\frac{\fixwidetilde{D}_{\textup{cp}}}{D} =
                \Bigg(
                \frac {      \tau \beta P_\textup{cp} \ln(\fixwidetilde{M})
                      }
                      {
                        \frac { 3 \Omega T_\textup{s} \ln(2)   } {\mu  |h|^2}
                        \big( \fixwidetilde{M}^{\frac{1}{2}} - 1 \big)^2
                        + P_\textup{o} T_\textup{s} \ln(2)
                      }
                \Bigg)^{\frac{1}{\beta+1}},
\end{equation}
\noindent and the optimal transmit power is given by
\begin{equation}\label{optimal-Pt-1}
   P^*_\textup{t} = \big( 1 - M^* \big) \frac { \Omega } { |h|^2},
\end{equation}
and the optimal compression ratio is given by
\begin{equation}\label{optimal-Dcp-1}
\frac{D^*_{\textup{cp}}}{D}=
    \begin{cases}
      \frac{\fixwidetilde{D}_{\textup{cp}}}{D},    &   \textup{if} \; \mathcal{Q}(\fixwidetilde{M},\fixwidetilde{D}_\textup{cp}) < T. \\
      \max  \big( \frac{D_\textup{min}}{D}, \frac{\fixwidehat{D}_{\textup{cp}}}{D} \big),        &   \textup{otherwise.}
    \end{cases}
\end{equation}
\noindent where $\frac{\fixwidehat{D}_{\textup{cp}}}{D} = \xi^{\frac{1}{\beta+1}}$ and $\xi$ is defined in \eqref{xi}.
\end{theorem}
\begin{IEEEproof}
The proof is provided in Appendix~\ref{B}.
\end{IEEEproof}

\vspace{0.1cm}
The insights from Theorem 1 are discussed in the following four remarks.

\begin{remark}
$\fixwidetilde{M}$ and $\fixwidetilde{D}_\textup{cp}$ provide a lower bound on the optimization problem in \eqref{opt-prob-p-1} for given instantaneous CGI, $|h|^2$. $\fixwidetilde{M}$ and $\fixwidetilde{D}_\textup{cp}$ are optimal design parameters when the first constraint in \eqref{opt-prob-p-1} is slack, i.e., $\mathcal{Q}(\fixwidetilde{M},\fixwidetilde{D}_\textup{cp}) < T$, and other constraints are also slack. On the other hand, $\fixwidehat{M}$ and $\fixwidehat{D}_\textup{cp}$ are optimal design parameters for optimization problem in \eqref{opt-prob-p-1} for given instantaneous CGI, $|h|^2$, when all constraints in \eqref{opt-prob-p-1} are slack except for the first constraint.
\end{remark}
\begin{remark}
In prior power-rate adaptation schemes \textup{\cite{UYSAL-2002, UYSAL-2004, Zafer-2008, Berry-2002, Zafer-2009, rao-2002}}, which do not consider data compression, the transmission policy is adapted with channel variations considering delay constraint under a fixed \textup{BER}. In these schemes, for a fixed amount of data to be transmitted, the transmission policy is adapted for the channel realization in each time block. However, in our case the transmitted compressed data size changes from one time block to the next, as a result of joint optimization of the transmission and compression policies adapting the channel realization in each time block.
\end{remark}
\begin{remark}
The classical works \textup{\cite{UYSAL-2002, UYSAL-2004}} have designed the transmission rate control policies which are adaptive to the channel conditions. These schemes do not consider data compression. Their optimal solution suggests that the energy cost of data communication is a strictly increasing function of the transmission rate. Therefore, the transmission rate should be minimized for the given delay bound, $T$, in order to minimize the energy cost of data communication cost. That is, the lowest transmission rate which will meet the delay constraint with equality is optimal for any value of the delay bound, $T$. However, in our case the combined data compression and transmission rate strategy suggests that there exists a lower bound on the total energy cost of compression and transmission. In this regard, the corresponding optimal design parameters cost a finite delay, $\mathcal{Q}$. Therefore, if the required delay constraint is larger than this delay only then these design parameters will maximize the lifetime and are optimal. Hence, in general, it is not optimal to transmit at the lowest transmission rate.
\end{remark}
\begin{remark}\label{remark-m}
The optimal constellation size given by \eqref{optimal-M-1} is real valued. Thus, for practical admissibility, the transmission policy should opt to select the closest value from the available set of modulation order values. If a lower value is closer then it can only be selected if the first constraint in \eqref{opt-prob-p-1} is slack, else a higher value should be selected which will surely satisfy the first constraint in \eqref{opt-prob-p-1}. This optimal practical value, denoted by $M^*_\textup{pr}$, is subsequently used to determine $P^*_\textup{t}$ and $\frac{D^*_{\textup{cp}}}{D}$. $M^*_\textup{pr}$ can be obtained using the following conditional expression
\begin{equation}\label{practical-m}
M^*_\textup{pr}=
    \begin{cases}
      \min  \big( 2^L, \nu_1  \big),    &
      \textup{if}\ |M^* - \nu_1| \leqslant |M^* - \nu_2| \: \textup{and} \: \mathcal{Q}( \nu_1, \breve{D}_{\textup{cp}}) < T.        \\
      \min  \big( 2^L, \nu_2 \big),    &   \textup{otherwise.}
    \end{cases}
\end{equation}
\noindent where $\min(\cdot)$ is the \textup{min} operation, $|\cdot|$ represents the absolute value, $2^L$ is the maximum modulation order supported by the system, $\nu_1 = 2^{\lfloor{\log_2 (M^*)}\rfloor}$,  $\nu_2 = 2^{\lceil{\log_2 (M^*)}\rceil}$, $\mathcal{Q}$ is defined in \eqref{optimal-Q-3}, $\lfloor{\cdot}\rfloor$ and $\left \lceil{\cdot} \right \rceil$ is the floor and ceil operations, respectively. Note, the compressed data size $\breve{D}_{\textup{cp}}$, is a function of the constellation size $2^{\lfloor{\log_2 (M^*)}\rfloor}$. %
\end{remark}

\subsubsection{Imperfect CGI availability}

The second problem we study considering imperfect CGI availability at the sensor node can be given as:

Problem~2: What is the optimal compression and transmission policy that minimizes the compression and transmission energy cost under specific delay and BER constraints, when quantized channel gain is known at the sensor node?

In Scenario~2, for a given channel realization $h$, the sink node feeds back the quantization level index $i$ to the sensor node, which implies CGI $|h|^2$, lies in the interval $[c_i,c_{i+1})$, i.e., $c_i~\leqslant~|h|^2~<~c_{i+1}$. Hence, there lies some channel uncertainty at the sensor node. As discussed above, the sensor node can minimize $\Psi$ in order to maximize the node-lifetime. Therefore, given $c_i~\leqslant~|h|^2~<~c_{i+1}$, the Problem~2 can be expressed as follows
\begin{equation}\label{opt-prob-p-2}
\begin{aligned}
& \underset{ M, P_{\textup{t}}, D_{\textup{cp}}} {\textup{minimize}}
& & \Psi (M, P_{\textup{t}}, D_{\textup{cp}})                           \\
& \textup{subject to}
& &   T_{\textup{cp}} +  T_{\textup{tx}} \leqslant T,  \quad \textup{BER}(M, P_{\textup{t}} ) \leqslant \phi, \\
& & & \forall \, c_i \leqslant |h|^2 < c_{i+1},  \quad P_{\textup{t}} \geqslant 0,  \quad M \geqslant 2,  \\
& & & M \leqslant M_\textup{max},   \quad  D_{\textup{cp}} \geqslant D_{\textup{min}}, \quad D_{\textup{cp}} \leqslant D.
\end{aligned}
\end{equation}
\noindent where the first constraint in \eqref{opt-prob-p-2} defines the delay constraint for the data delivery, thus both compression and transmission processes should be completed within the deadline, and the second constraint in \eqref{opt-prob-p-2} mandates that the $\textup{BER}$ should be below or equal to a specific threshold value denoted by $\phi$, and the third constraint in \eqref{opt-prob-p-2} represents the condition that the actual CGI, $|h|^2$, lies within the interval $[c_i,c_{i+1})$.

Accordingly, the solution to the problem in \eqref{opt-prob-p-2} is given by the following corollary.

\begin{corollary}
In solving the optimization problem in \eqref{opt-prob-p-2}, the optimal constellation size, transmit power, and compression ratio can be obtained using \eqref{optimal-M-1}, \eqref{optimal-Pt-1}, and \eqref{optimal-Dcp-1}, respectively, by replacing $|h|^2$ with $c_i$.
\end{corollary}

\begin{IEEEproof}
The BER, defined in \eqref{ber}, is a function of constellation size, $M$, and the CGI, $|h|^2$. Recall, the sink node feeds back the index $i$ corresponding to the quantization interval, where the actual CGI lies in the interval $c_i~\leqslant~|h|^2~<~c_{i+1}$, as mentioned earlier in Section~\ref{sec-CGI}. In order to meet the required BER constraint, defined in \eqref{opt-prob-p-2}, the sensor node selects the smallest value, i.e., $c_i$, in the quantization interval $[c_i,c_{i+1})$ indicated by the index $i$. This underestimation ensures the quantized CGI, $c_i$, is always less than or equal to the actual CGI, i.e., $c_i \leqslant |h|^2$. Hence the required BER constraint, defined in \eqref{opt-prob-p-2}, is naturally satisfied for any given level of channel uncertainty due to CGI quantization, i.e., $\textup{BER}(c_i)~\leqslant~\textup{BER}(|h|^2)~\leqslant~\phi$. Therefore, by replacing $|h|^2$ with $c_i$ in \eqref{ber} and following similar steps provided in Appendix~\ref{A}, the optimal constellation size, transmit power, and compression ratio can be obtained, given $c_i \leqslant~|h|^2 < c_{i+1}$.
\end{IEEEproof}
\vspace{-0.2cm}
\begin{remark}
The values for the optimal design parameters for optimization problem defined in \eqref{opt-prob-p-2} can be obtained offline using Corollary~1. This is because there are only a finite set of quantization levels, each corresponding to a set of design parameters values. For example, if 5 feedback bits are used then only 32 different values for the optimal design parameters need to be computed. These values can then be used in a given time block based on the received quantized CGI feedback value, denoted by $c_i$. Hence, the proposed solution for Scenario~2 is practical and feasible for energy constrained wireless sensor based MTC devices.
\end{remark}

\subsection{Statistical CGI availability}

The third problem we study considering availability of only statistical information of CGI at the sensor node can be summarized as follows:

Problem~3: What is the optimal compression and transmission policy that minimizes the compression and transmission energy cost under specific delay and BER constraints, when only statistical information about the channel gain is known at the sensor node?

Since, the instantaneous CGI is not available, thus the design parameters cannot be adapted to any given channel condition and cannot guarantee a certain BER, in a given time block. However, we can determine the optimal design parameters which are set to be the same for each time block with an objective to minimize the compression and transmission energy cost,~$\Psi$. Note that $\mathbb{E}[\Psi]=\Psi$ when the design parameters are set to be the same for all time blocks. Given the fading power gain distribution, $f(|h|^2)$, Problem~3 can be expressed as follows
\begin{equation}\label{opt-prob-p-3}
\begin{aligned}
& \underset{ M, P_{\textup{t}}, D_{\textup{cp}}} {\textup{minimize}}
& & \Psi (M, P_{\textup{t}}, D_{\textup{cp}})                           \\
& \textup{subject to}
& &   T_{\textup{cp}} +  T_{\textup{tx}} \leqslant T,   \quad  \mathbb{P}\{\textup{BER} \leqslant \phi \} \geqslant \vartheta,  \\
& & & P_{\textup{t}} \geqslant 0, \quad M \geqslant 2,    \quad M \leqslant M_\textup{max},  \\
& & &  D_{\textup{cp}} \geqslant D_{\textup{min}}, \quad D_{\textup{cp}} \leqslant D.
\end{aligned}
\end{equation}
\noindent where the first constraint in \eqref{opt-prob-p-3} defines the delay constraint for the data delivery, thus both compression and transmission processes should be completed within the deadline, and the second constraint in \eqref{opt-prob-p-3} mandates that the probability of having as acceptable level of BER should be greater than certain percentage. Specifically,  $\phi$ denote the maximum acceptable BER and $\vartheta$ denote the required minimum probability of achieving the acceptable BER performance. The third constraint in \eqref{opt-prob-p-3} guarantees a given BER performance with a certain probability in each time block. Note that an alternative way of constraining the BER performance is to put an upper bound on the average BER over all time blocks. But we do not adopt it because it gives minimal control over the BER performance in each time block.

\begin{proposition}\label{propo-3}
The optimal $P_{\textup{t}}$ to minimize $\Psi(M, P_{\textup{t}}, D_{\textup{cp}})$ for given values of $M$ and $D_{\textup{cp}}$ while satisfying the constraints in \eqref{opt-prob-p-3} is given by
\begin{equation}\label{min-Pt-propo-3}
   P_\textup{t} = (M - 1) \frac {\Omega} {\varsigma \ln ( \vartheta)}.
\end{equation}
\end{proposition}
\begin{IEEEproof}
The proof is provided in Appendix~\ref{C}.
\end{IEEEproof}

Using the result in Proposition 2, substituting $T_{\textup{cp}}$, $T_{\textup{tx}}$, $r$, $P_\textup{tx}$, $\varepsilon$ and $P_{\textup{t}}$ from \eqref{t-comp}, \eqref{t-comm}, \eqref{r}, \eqref{p-comm}, \eqref{mqam-par} and  \eqref{min-Pt-propo-3}, respectively, in \eqref{psi} yields $\Psi$ as a function of $M$ and $D_{\textup{cp}}$ as follows
\begin{equation}\label{psi-dc-3}
 \Psi(M,D_{\textup{cp}})  =  \tau D^{\beta+1} D_{\textup{cp}}^{-\beta} P_{\textup{cp}} - \tau D P_{\textup{cp}}
                         +  \frac{ D_{\textup{cp}} T_\textup{s} \ln(2) }{  \ln(M) }  \Bigg(  \frac {3 \Omega ({M}^{\frac{1}{2}} - 1)^2 }  {\mu \varsigma  \ln ( \vartheta )} + P_\textup{o} \Bigg).
\end{equation}

Now a simpler equivalent optimization problem with only two design parameters, i.e., $M, D_{\textup{cp}}$, needs to be solved and the third parameter $P_\textup{t}$ can be obtained using the result in Proposition~2. Accordingly, the solution to the optimization problem in \eqref{opt-prob-p-3} is given by the following theorem.

\begin{theorem}
In solving the optimization problem in \eqref{opt-prob-p-3}, the optimal constellation size is given by the following conditional expression
\begin{equation}\label{optimal-M-3}
M^*=
    \begin{cases}
      \fixwidetilde{M},    &   \textup{if} \; \mathcal{Q}(\fixwidetilde{M},\fixwidetilde{D}_\textup{cp}) < T. \\
      \min  \big( \fixwidehat{M}, M_\textup{max} \big),      &   \textup{otherwise.}
    \end{cases}
\end{equation}
\noindent where $\fixwidetilde{M}$ and $\fixwidehat{M}$ are given by the solution of the following equations which can be solved numerically using \textup{VPA} method
\begin{equation}\label{optimal-M-tilde-3}
 - \frac { 3 \Omega } {\mu \varsigma \ln (\vartheta)}
  \big( \fixwidetilde{M}^{\frac{1}{2}} - 1 \big) \big( (\ln(\fixwidetilde{M})-1) \fixwidetilde{M}^{\frac{1}{2}} + 1 \big)  + P_\textup{o}  =  0,
\end{equation}
\begin{equation}\label{optimal-M-bar-3}
\frac{T}{D} + \tau - \tau
                       \zeta^{\frac{-\beta}{\beta+1}}
  =  \frac{ T_{\textup{s}} \ln(2) } {\ln(\fixwidehat{M})}
                     \zeta^{\frac{1}{\beta+1}},
\end{equation}
\noindent where
\begin{equation}\label{zeta}
\zeta =         \frac {
                             P_\textup{cp}  {-}   P_\textup{o}
                             {+} \frac { 3 \Omega  } {\mu \varsigma \ln (\vartheta)}  \big( \fixwidehat{M}^{\frac{1}{2}} {-} 1 \big) \big( (\ln(\fixwidehat{M}){-}1) \fixwidehat{M}^{\frac{1}{2}} {+} 1 \big)
                      }
                      {
                        - \frac { 3 \Omega  } {\varsigma \tau \beta \mu  \ln (\vartheta)}
                        \big( \fixwidehat{M}^{\frac{1}{2}} - \fixwidehat{M} \big)
                      },
\end{equation}
\noindent respectively,
\begin{equation}\label{optimal-Q-3}
\mathcal{Q}(\fixwidetilde{M},\fixwidetilde{D}_\textup{cp}) \triangleq \tau D^{\beta+1} \fixwidetilde{D}_{\textup{cp}}^{-\beta} - \tau D +  \frac{ \fixwidetilde{D}_{\textup{cp}} T_\textup{s} }{  \log_2 \big( \fixwidetilde{M} \big) },
\end{equation}
\noindent and
\begin{equation}\label{optimal-Dcp-tilde-3}
\frac{\fixwidetilde{D}_{\textup{cp}}}{D} =
                \Bigg(
                \frac {      \tau \beta P_\textup{cp} \ln(\fixwidetilde{M})
                      }
                      {
                        - \frac { 3 \Omega T_\textup{s} \ln(2)   } {\mu \varsigma  \ln (\vartheta)}
                        \big( \fixwidetilde{M}^{\frac{1}{2}} - 1 \big)^2
                        + P_\textup{o} T_\textup{s} \ln(2)
                      }
                \Bigg)^{\frac{1}{\beta+1}},
\end{equation}
\noindent and the optimal transmit power is given by
\begin{equation}\label{optimal-Pt-3}
   P^*_\textup{t} = \big( M^* - 1 \big) \frac { \Omega } {\varsigma \ln (\vartheta)},
\end{equation}
and the optimal compression ratio is given by
\begin{equation}\label{optimal-Dcp-3}
\frac{D^*_{\textup{cp}}}{D}=
    \begin{cases}
      \frac{\fixwidetilde{D}_{\textup{cp}}}{D},      &   \textup{if} \; \mathcal{Q}(\fixwidetilde{M},\fixwidetilde{D}_\textup{cp}) < T. \\
      \max  \big( \frac{D_\textup{min}}{D}, \frac{\fixwidehat{D}_{\textup{cp}}}{D} \big),        &   \textup{otherwise.}
    \end{cases}
\end{equation}
\noindent where $\frac{\fixwidehat{D}_{\textup{cp}}}{D} = \zeta^{\frac{1}{\beta+1}}$ and $\zeta$ is defined in \eqref{zeta}.
\end{theorem}
\begin{IEEEproof}
The proof follows similar steps as the proof of Theorem 1. Hence, it is omitted for brevity.
\end{IEEEproof}

\begin{remark}
As discussed earlier in \textup{Remark \ref{remark-m}}, for practical admissibility the optimal constellation size, $M^*$, needs to be scaled to the closest achievable practical value using \eqref{practical-m}. Similarly, $P^*_\textup{t}$ needs to meet the maximum power bound of the battery.
\end{remark}
\begin{remark}
The values for the optimal design parameters for optimization problem defined in \eqref{opt-prob-p-3} can be obtained offline using Theorem~2. It is because these values are fixed for all time blocks. Thus, the proposed solution for Scenario~3 is practical and feasible for energy constrained wireless sensor based MTC devices.
\end{remark}
\begin{remark}
An alternative scenario to consider is to use an average BER constraint even when the instantaneous CGI is available. In this case, the lifetime can be further extended by dropping the sensed data in a given time block in which the channel is in deep fade. Let us replace the BER constraint function in \eqref{opt-prob-p-1} as follows: $\mathbb{P}\{\textup{BER} \leqslant \phi \} \geqslant \vartheta$. Thus, the probabilistic BER constraint can directly give a threshold on channel gain, given by $\Theta=- \varsigma \ln \big( \vartheta \big)$, above which data transmission should occur. Then the sensor only employs Theorem~1 to design compression and transmission when the instantaneous channel gain is above the threshold,~$\Theta$.
\end{remark}

\section{Results}\label{sec-results}

In this section, we present the numerical results to illustrate the performance of the proposed Scenarios 1-3 jointly optimizing data compression and transmission rate. Unless specified otherwise, the values for the system parameters shown in Table~\ref{para-table} are adopted.

To illustrate the advantage of joint optimization of data compression and transmission rate, we are interested in prior works which provide solutions for the data transmission design to maximize node-lifetime. To the best of our knowledge, the prior works \cite{rate, UYSAL-2004, Zafer-2008} are the most relevant and most recent schemes which can be compared to our proposed scheme. In this regard, we adopt the data transmission design policies proposed by these schemes for our considered system model except that data compression is not employed. Moreover, when our considered bit error rate and delay constraints are applied, the resultant data transmission design problem can be given as in (17) by substituting $D_{\textup{cp}} = D$. We refer to this adapted scheme as the baseline scheme. The strategy followed to optimize the transmission rate policy for the baseline scheme is essentially the same as in the state of the art \cite{rate, UYSAL-2004, Zafer-2008}.
The optimal constellation size for this scheme, denoted by $M^*_\textup{nc}$, can be obtained using Proposition~3.

\begin{proposition}
The optimal constellation size to maximize lifetime without performing compression while satisfying constraints given in \eqref{opt-prob-p-1} is given by the following conditional expression
\begin{equation}\label{no-compression-M}
M^*_\textup{nc}=
    \begin{cases}
      \fixwidetilde{M},    &   \textup{if} \; \frac{ D T_\textup{s} }{  \log_2 \big( \fixwidetilde{M} \big) } < T. \\
      \exp \Big( \frac{D T_\textup{s} \ln (2)} {T} \Big),      &   \textup{otherwise.}
    \end{cases}
\end{equation}
\noindent where $\fixwidetilde{M}$ is given by the solution of the following equation which can be solved numerically using the \textup{VPA} method
\begin{equation}\label{no-compression-M-tilde}
 \frac { 3 \Omega  } {\mu  |h|^2}
  \big( \fixwidetilde{M}^{\frac{1}{2}} - 1 \big) \big( (\ln(\fixwidetilde{M})-1) \fixwidetilde{M}^{\frac{1}{2}} + 1 \big)  + P_\textup{o}  =  0,
\end{equation}
\noindent and the optimal transmit power is given by
\begin{equation}\label{no-compression-M-Pt}
   P^*_\textup{t} = \big( 1 - M^*_\textup{nc} \big) \frac { \Omega } {|h|^2}.
\end{equation}
\end{proposition}
\begin{IEEEproof}
The proof follows similar steps as the proof of Theorem 1 and substituting $D_{\textup{cp}} = D$. Hence, it is omitted for brevity.
\end{IEEEproof}

\begin{table*}[]
\centering
\caption{System Parameter Values.}
\label{para-table}
\begin{tabular}{|l|c|c||l|c|c|} \hline
\textbf{Name}                                & \textbf{Symbol}  & \textbf{Value}    & \textbf{Name}    & \textbf{Symbol}  & \textbf{Value}            \\ \hline
Drain efficiency of power amplifier          & $\mu$            & 0.35              & Operating voltage& $V_\textup{op}$  & 3 V                       \\
Scale parameter for CGI's $pdf$              & $\varsigma$      & 1                 & Battery capacity & $B_\textup{cap}$ & 9000 As                   \\
Power cost of compression                    & $P_\textup{cp}$  & 24 mW             & Symbol period    & $T_\textup{s}$   & 16 $\mu$s                 \\
Power cost of synthesizer                    & $P_\textup{syn}$ & 50 mW             & constant         & $\omega_2$       & 0.2                       \\
Power cost of filter                         & $P_\textup{fil}$ & 2.5 mW            & constant         & $\omega_1$       & 1.5                       \\
Power cost of mixer                          & $P_\textup{mix}$ & 30.3 mW           & distance         & $d$              & 20 m                      \\
Uncompressed data                            & $D$              & 20kb             & Noise power      & $\sigma^2$       & -174 dBm                  \\
Per bit processing time                      & $\tau$           & 0.35 ns/b         & BER constraint   & $\phi$           & $10^{-3}$                 \\
Compression cost parameter                   & $\beta$          & 5                 & Delay constraint & $T$              & 50 ms                     \\ \hline
\end{tabular}
\end{table*}

\subsection{Advantage of Proposed Scheme}

\begin{figure}
\begin{subfigure}{.5\textwidth}
\centering
\includegraphics[scale=1]{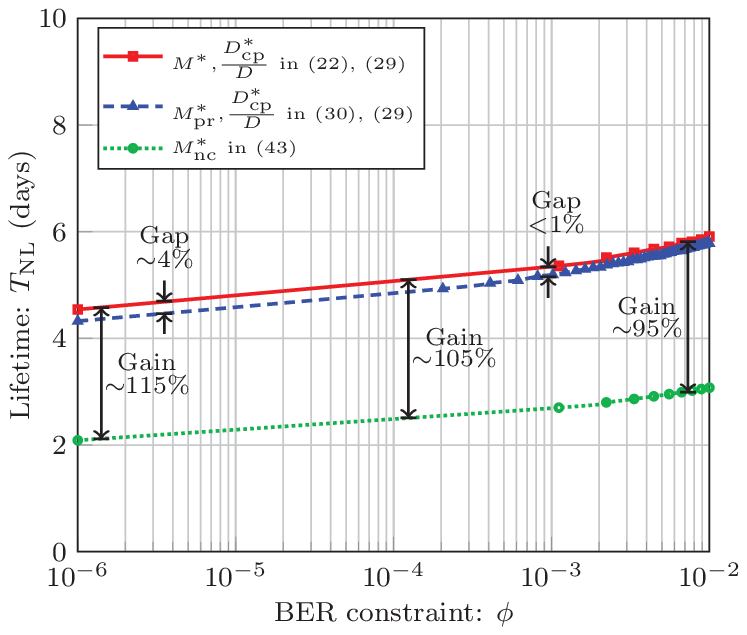}
\caption{} \label{nm-validation-BER}
\end{subfigure}
\begin{subfigure}{.5\textwidth}
\centering
\includegraphics[scale=1]{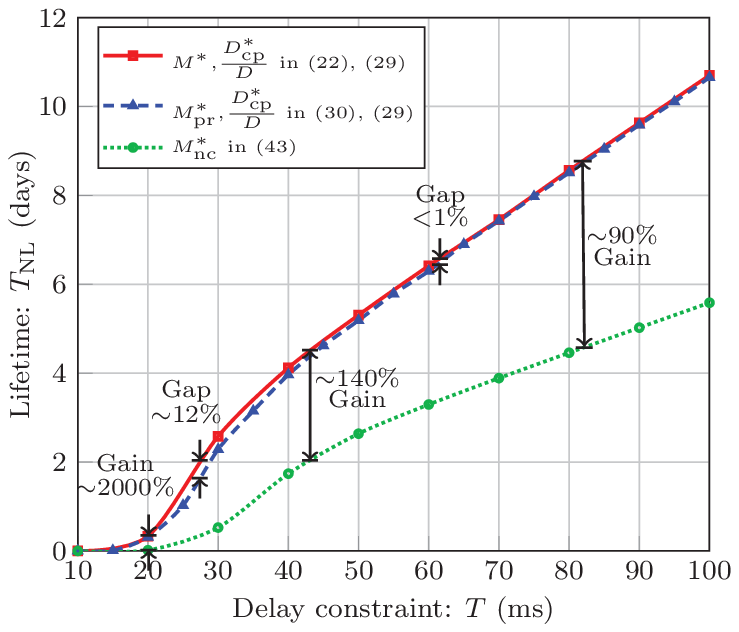}
\caption{} \label{nm-validation-T}
\end{subfigure}
\caption{Lifetime vs. BER (a) and delay constraints (b) for Scenario 1 with and without data compression, when $T$ = 50 ms for (a) and $\phi$ = $10^{-3}$  for (b).}
\end{figure}

Fig. \ref{nm-validation-BER} plots the node-lifetime, $T_\textup{NL}$ (days), versus the BER constraint, $\phi$, for Scenario~1 and system parameters in Table~\ref{para-table}. The lifetime is plotted with the optimal (real valued) $M^*$ in \eqref{optimal-M-1}, the practical (quantized value) ${M_\textup{pr}}^*$ in \eqref{practical-m} and the baseline scheme in ${M_\textup{nc}}^*$ in \eqref{no-compression-M} in Fig.~\ref{nm-validation-BER}. We can see that the gain compared to the baseline scheme is significant - between 95\% to 115\% for the considered range of BER constraint. This shows the advantage of joint optimal compression and transmission rate control. \emph{In addition, we can see that the node-lifetime is not so significantly affected by the BER constraint.} As the BER constraint is relaxed, the lifetime slightly increases. For instance, as BER constraint is varied from stringent BER requirement, i.e., $10^{-6}$, to loose BER requirement, i.e., $10^{-2}$, the lifetime only changes by around 30\%. Finally, the performance with practical modulation scheme is very close to the optimal performance, e.g., at $\phi=10^{-5}$ the gap is less than 4\%.

Fig. \ref{nm-validation-T} plots the node-lifetime, $T_\textup{NL}$ (days), versus the delay constraint, $T$ (ms), for Scenario~1 and system parameters in Table~\ref{para-table}. The node-lifetime is plotted with the optimal (real valued) $M^*$ in \eqref{optimal-M-1}, the practical (quantized value) ${M_\textup{pr}}^*$ in \eqref{practical-m} and the baseline scheme in ${M_\textup{nc}}^*$ in \eqref{no-compression-M} in Fig.~\ref{nm-validation-T}. The compression ratio and transmission rate are optimized through $M$ and $D_\textup{cp}$. Though inherently independent, jointly optimizing $M$, $P_\textup{t}$, and $D_\textup{cp}$ to maximize the node-lifetime yields their relationship as follows. As given in Theorem~1, the optimal constellation size has a direct relationship with both the optimal transmit power level and the optimal compressed data size. Note, a smaller value of compressed data size means a high level of compression is applied.

The explanation of the gains shown in Fig.~\ref{nm-validation-T} is as follows. The performance gains are comparatively lower ($\backsim90$\%) in the case when the delay bound is loose ($\backsim80 ms$). This is because the data communication energy cost without employing compression almost increases exponentially with the employed transmission rate. When the delay bound is stringent then relatively lower transmission rate is good enough to transmit data whilst satisfying the BER and delay constraints. Since, the data communication energy cost without employing compression is not too significant, employing compression improves the lifetime with relatively lesser magnitude. The performance gains are huge ($\backsim2000$\%) when the delay bound is stringent ($\backsim20 ms$). This is because, if the delay bound is stringent then a very high transmission rate is required to transmit the given data. Thus, the per bit transmission energy cost is very large for a stringent delay bound. However, in the proposed schemes, an optimal level of compression is applied to the data and the energy of applying (optimal level of) compression is much less than transmitting data at a very high transmission rate, i.e., per bit compression energy cost is very low. Therefore, with the help of data compression the amount of data is reduced and in addition the compressed data is transmitted at a comparatively lower transmission rate. As a result large gains are possible when optimal level of compression and transmission rate policy is employed.

From Fig. \ref{nm-validation-BER} and \ref{nm-validation-T}, we can say that joint optimization is much better than no compression under any BER and delay constraints. In addition, the performance gain observed ranges from $90\%$ to $2000\%$ and is most profound when the delay constraint is stringent, which demonstrates the suitability of applying the proposed scheme in the low latency regime.

\subsection{Impact of BER and Delay Constraints - Scenario 1}

\begin{figure}
\begin{subfigure}{.5\textwidth}
\centering
\includegraphics[scale=1]{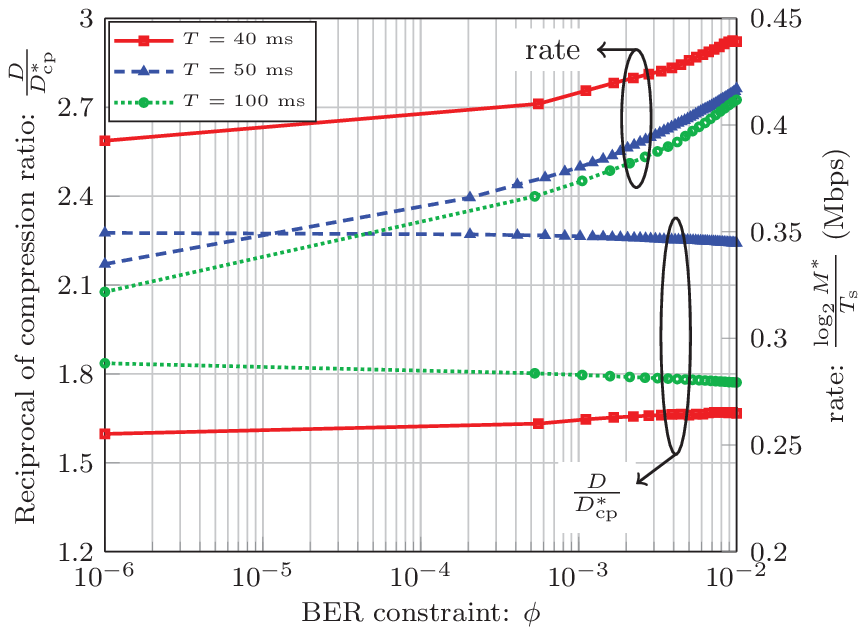}
        \vspace*{-2mm}
\caption{} \label{nm-sch1-BER}
\end{subfigure}
\begin{subfigure}{.5\textwidth}
\centering
\includegraphics[scale=1]{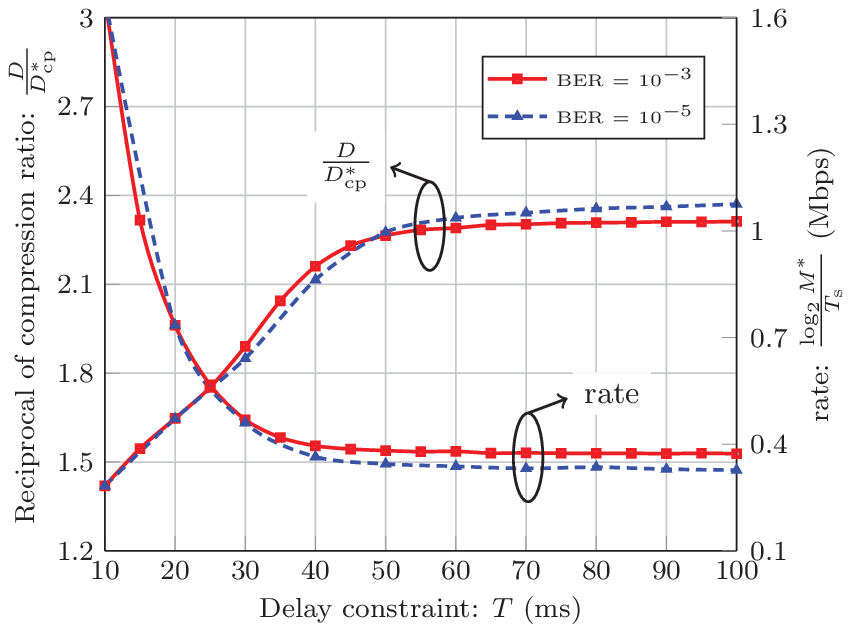}
        \vspace*{-2mm}
\caption{} \label{nm-sch1-T}
\end{subfigure}
\caption{Optimal compression ratio and transmission rate vs. BER (a) and delay constraints (b) for Scenario 1, when $T$ = $\{40,50,100\}$ ms for (a) and $\phi$ = $\{10^{-3},10^{-5}\}$ for (b).}
\end{figure}

Fig. \ref{nm-sch1-BER} plots the reciprocal of the compression ratio and the transmission rate vs. the BER constraint for different delay constraint values, for Scenario~1. \emph{Note, high compression ratio implies less compression is applied, thus we plot its reciprocal for better clarity}. The performance observed in Fig. \ref{nm-validation-BER} is explained in terms of design parameters in Fig. \ref{nm-sch1-BER}. We can see that as the BER constraint becomes less stringent, the transmission rate increases, especially in the range $10^{-3}$ to $10^{-2}$, whereas the level of compression remains almost constant. Hence, the optimal level of compression is insensitive to the change in the BER requirement. However, the optimal transmission rate increases as BER constraint is relaxed. \emph{Thus, the best choice is to keep a constant optimal level of compression and adapt transmission rate as per the BER requirement.}

Fig. \ref{nm-sch1-T} plots the reciprocal of the compression ratio and the transmission rate vs. the delay constraint for different BER constraint values, for Scenario~1. The performance observed in Fig.~\ref{nm-validation-T} is explained in terms of design parameters in Fig. \ref{nm-sch1-T}. We can see that as the delay constraint becomes less stringent, the transmission rate significantly increases and the level compression also significantly increases until the upper bound $\mathcal{Q}$ in \eqref{optimal-Q-1} is reached. Beyond that, both design parameters remains almost constant.

Overall, it is best to reduce compression and increase the transmission rate when the delay constraint gets stringent and vice versa. The optimal design of compression and transmission
rate is more sensitive to the delay constraint when the system requires low latency.

\subsection{Effect of no. of feedback bits - Scenario 2}

\begin{figure}
\centering
\includegraphics[scale=1]{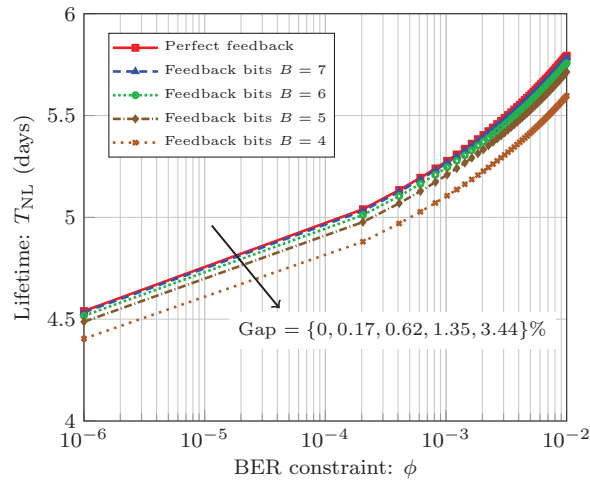}
\caption{Lifetime vs. BER constraint for Scenario 2 for different level of channel uncertainty, when $T$~=~50~ms.}
\label{nm-sch2-bits}
\end{figure}

Fig. \ref{nm-sch2-bits} plots the lifetime versus the BER constraint with perfect feedback for Scenario~1 and quantized feedback for Scenario~2. The gap between the performance of Scenario~1 (perfect CGI availability at the sensor node) and Scenario~2 (quantized CGI availability at the sensor node) is small for practical number of feedback bits. It can be observed that a relatively small number of feedback bits are enough, e.g., 6 bits achieve within 0.6\% optimal performance.

\subsection{Impact of BER and Delay Constraints - Scenario 3}

\begin{figure}
\begin{subfigure}{.5\textwidth}
\centering
\includegraphics[scale=1]{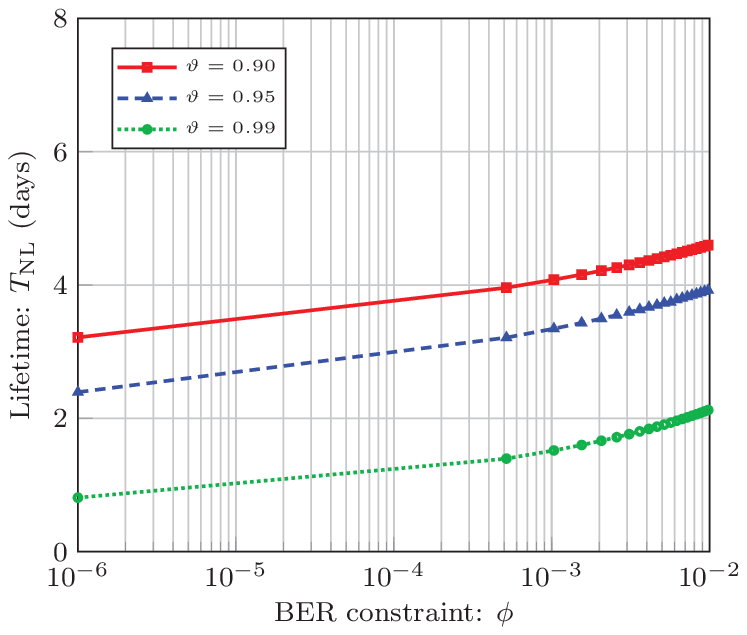}
\caption{} \label{nm-sch3-vartheta}
\end{subfigure}
\begin{subfigure}{.5\textwidth}
\centering
\includegraphics[scale=1]{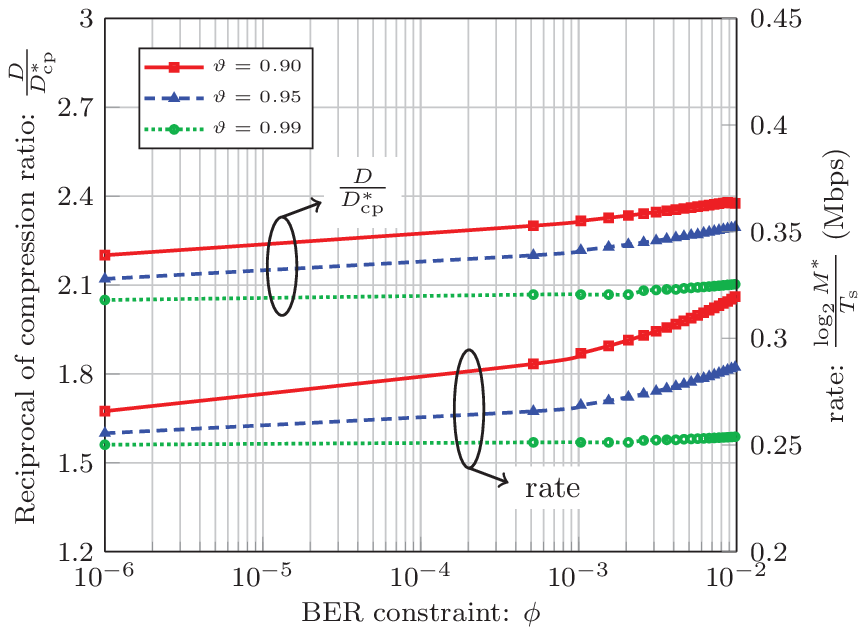}
\caption{} \label{nm-sch3-BER}
\end{subfigure}
\begin{subfigure}{.5\textwidth}
\centering
\includegraphics[scale=1]{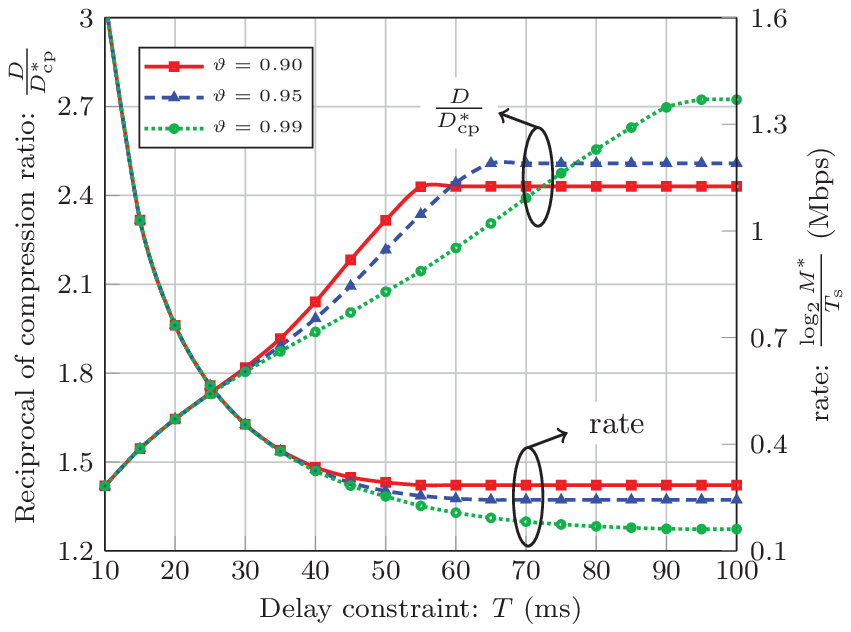}
\caption{} \label{nm-sch3-T}
\end{subfigure}
\caption{Optimal compression and transmission rate vs. BER (a) and delay constraints (b) for Scenario 3 for different probabilistic BER performance requirement, when $T$~=~50~ms for (a) and (b) and $\phi = 10^{-3}$ for (c).}
\end{figure}

Fig. \ref{nm-sch3-vartheta} plots the lifetime versus the BER constraint for different level of probabilistic BER performance requirement, $\vartheta$, for Scenario~3. The lifetime is impacted by the BER constraint in a similar manner as compared to Scenario~1 shown in Fig. \ref{nm-validation-BER}. The lifetime non-linearly decreases with increase in the BER performance requirement.

Fig. \ref{nm-sch3-BER} shows that for a given BER constraint, $\phi$, as $\vartheta$ increases both the transmission rate and the level of compression decrease. However, in the case of stringent BER requirement, i.e., $\vartheta~=~0.99$, both the level of compression and transmission rate remains almost constant for different values of $\phi$. For a given value of $\vartheta$, the level of compression increases with $\phi$, unlike the trend observed in Fig. \ref{nm-sch1-BER} for Scenario~1.

Fig. \ref{nm-sch3-T} shows that for a given delay constraint, as $\vartheta$ increases, the transmission rate decreases. The level of compression displays different trend as compared to Scenario~1 in Fig. \ref{nm-validation-T}. This is because the value of the upper bound $\mathcal{Q}$ in \eqref{optimal-Q-3} increases as $\vartheta$ increases. Thus, the level of compression increase until this upper bound is reached and afterwards it remains almost constant.

\section{Conclusion}

In this work, we investigated the joint optimization of compression and transmission strategy for an energy-constrained sensor node, and illustrated their tradeoff. We showed that the joint optimization performs much better than only optimizing transmission without compression under any BER and delay constraints. The performance gain observed ranges from $90\%$ to $2000\%$ and is most profound when the delay constraint is stringent. Overall, it is best to reduce compression and increase the transmission rate when the delay constraint gets more stringent and vice versa. The optimal level of compression is insensitive to the change in the BER requirement. However, the optimal transmission rate increases as BER constraint is relaxed. Hence, the best choice is to keep a constant optimal level of compression and adapt transmission rate as per the BER requirement. The optimal level of compression has an inverse relationship with severity of the BER requirement when the delay constraint is stringent and vice versa. In this paper, we have assumed that new data is generated once the previous one is processed. In some event monitoring applications, the sensed data can arrive before the previous data is completely transmitted. Future work can consider transmission and compression design for such a scenario.

\appendices

\section{Proof of Proposition 1}\label{A}

Substituting  BER constraint \eqref{ber} and SNR expression \eqref{snr} in second constraint in \eqref{opt-prob-p-1} yields
\begin{equation}\label{ber1-3}
  \omega_2 \; \textup{exp} \bigg( -  \frac { \omega_1 } {(M-1)} \frac {\kappa P_\textup{t} |h|^2}{\sigma^2 d^\alpha}  \bigg) \leqslant \phi.
\end{equation}

For given instantaneous CGI, $|h|^2$, a lower bound on the transmit power, $P_{\textup{t}}$, can be obtained by rearranging \eqref{ber1-3} as follows
\begin{equation}\label{min-Pt-0}
   P_\textup{t} \geqslant (1 - M) \frac { \sigma^2 d^\alpha \ln\big( \phi/\omega_2 )} {\omega_1 \kappa |h|^2}.
\end{equation}

By substituting $T_{\textup{cp}}$, $T_{\textup{tx}}$, $P_\textup{tx}$, $r$ and $\varepsilon$ from \eqref{t-comp}, \eqref{t-comm}, \eqref{p-comm}, \eqref{r} and \eqref{mqam-par}, respectively, in \eqref{psi} yields $\Psi$ as a function of $M$, $D_{\textup{cp}}$ and $P_{\textup{t}}$, which can be expressed as follows
\begin{equation}\label{psi-dc-1-0}
  \Psi  =    \tau D^{\beta+1} D_{\textup{cp}}^{-\beta} P_{\textup{cp}} - \tau D P_{\textup{cp}}
                    +  \frac{ D_{\textup{cp}} T_\textup{s} \ln(2) }{  \ln(M) }
                       \bigg(  \frac{3 P_\textup{t} ({M}^{\frac{1}{2}}-1)}{\mu ({M}^{\frac{1}{2}}+1)}   + P_\textup{o} \bigg).
\end{equation}

For given values of $M$ and $D_{\textup{cp}}$, $\Psi$ is an increasing function of $P_{\textup{t}}$. Hence, the best choice of $P_{\textup{t}}$ to minimize $\Psi$ while satisfying the constraint in \eqref{min-Pt-0} is the minimum value obtained by setting \eqref{min-Pt-0} with equality. Thus, $P_{\textup{t}}$ can be expressed as a function of $M$, as given in \eqref{min-Pt-propo-1}.

\section{Proof of Theorem 1}\label{B}

It can be shown that \eqref{psi-dc-1} is not convex in $M$. By substitution of variable $M = \textup{exp}(z)$ in \eqref{psi-dc-1}, $\fixwidetilde{\Psi}$ can be equivalently defined as
\begin{equation}\label{psi-dc-2}
  \fixwidetilde{\Psi}(z,D_{\textup{cp}})  =     \tau D^{\beta+1} D_{\textup{cp}}^{-\beta} P_{\textup{cp}} - \tau D P_{\textup{cp}}
                     -  \frac{  T_\textup{s} \ln(2) }{ z D_{\textup{cp}}^{-1} }
                       \bigg(   \frac {3 \Omega (\textup{exp}(z/2) - 1)^2  }  {\mu  |h|^2} - P_\textup{o} \bigg).
\end{equation}

Accordingly, the problem defined in \eqref{opt-prob-p-1} can be equivalently given as follows
\begin{equation}\label{opt-prob-1-transformed}
\begin{aligned}
& \underset{ z, D_{\textup{cp}}} {\textup{minimize}}
& &   \fixwidetilde{\Psi}(z,D_{\textup{cp}}) \\
& \textup{subject to}
& & \tau D^{\beta+1} D_{\textup{cp}}^{-\beta} - \tau D + z^{-1}D_{\textup{cp}} T_\textup{s} \ln(2) - T \leqslant 0, \\
& & & 2 - \exp(z) \leqslant 0, \quad \exp(z) - \ln(M_\textup{max}) \leqslant 0, \\
& & & D_{\textup{min}}-D_{\textup{cp}} \leqslant 0, \quad D_{\textup{cp}} - D \leqslant 0.
\end{aligned}
\end{equation}

For brevity we omit the proof, however using basic calculus and with some algebraic manipulation, it can be shown that the problem in \eqref{opt-prob-1-transformed} is a convex optimization problem. Lagrangian function for \eqref{opt-prob-1-transformed} can be given as in \eqref{lag-main-1}, where $\Lambda_i \in \mathbf{\Lambda}=\{\Lambda_1, \Lambda_2, \Lambda_3, \Lambda_4, \Lambda_5\}$ is the Lagrangian multiplier associated with the $i$th constraint.

\begin{equation}\label{lag-main-1}
\begin{aligned}
 \mathcal{L}(z,D_{\textup{cp}},\mathbf{\Lambda})  \,\,\, = \,\,\, &
 \tau D P_{\textup{cp}} \Big( \frac{D^{\beta}}{D_{\textup{cp}}^{\beta}} {-} 1  \Big)
                                 -  \frac{ T_\textup{s} \ln(2) }{ z D_{\textup{cp}}^{-1} }
                                \bigg(   \frac {3 \Omega (\textup{exp}(z/2) {-} 1)^2 }  {\mu |h|^2} {-} P_\textup{o} \bigg)
                                \\
                                & + \Lambda_{1}
                                \bigg(
                                 \tau D \Big( \frac{D^{\beta}}{D_{\textup{cp}}^{\beta}}{-}1  \Big)
                                {+}  \frac{  T_\textup{s} \ln(2) }{ z D_{\textup{cp}}^{-1} } {-} T \bigg)
                                + \Lambda_{2} \big(2-\textup{exp}(z)\big)
                                \\
                                &
                                + \Lambda_{3} \big(\exp(z) - \ln(M_\textup{max})\big)
                                + \Lambda_{4} ( D_{\textup{min}}-D_{\textup{cp}})
                                + \Lambda_{5} ( D_{\textup{cp}} - D),
\end{aligned}
    \end{equation}

The Karush-Kuhn-Tucker (KKT) conditions for \eqref{opt-prob-1-transformed} are:
\begin{subequations}
\begin{equation}
\begin{aligned}
& \tau D^{\beta+1} D_{\textup{cp}}^{-\beta} - \tau D  +  z^{-1} D_{\textup{cp}} T_\textup{s} \ln(2) - T \leqslant 0,  \\
& 2-\textup{exp}(z) \leqslant 0, \quad \exp(z) - \ln(M_\textup{max}) \leqslant 0, \\
& D_{\textup{min}} - D_{\textup{cp}} \leqslant 0, \quad D_{\textup{cp}} - D \leqslant 0,
\end{aligned}
\end{equation}
\begin{equation}
\begin{aligned}
& \Lambda_{1} \geqslant 0, \,\,\,  \Lambda_{2} \geqslant 0,  \,\,\, \Lambda_{3} \geqslant 0, \,\,\, \Lambda_{4} \geqslant 0, \,\,\, \Lambda_{5} \geqslant 0,
\end{aligned}
\end{equation}
\begin{equation}\label{3rd-kkt-1}
\begin{aligned}
& \Lambda_{1}  \Big( \tau D^{\beta+1} D_{\textup{cp}}^{-\beta} - \tau D  +  z^{-1} D_{\textup{cp}} T_\textup{s} \ln(2) - T \Big) = 0, \\
& \Lambda_{2} (2-\textup{exp}(z)) = 0, \quad \Lambda_{3} (\exp(z) - \ln(M_\textup{max})) = 0, \\
& \Lambda_{4} (D_{\textup{min}} - D_{\textup{cp}}) = 0,  \quad \Lambda_{5} ( D_{\textup{cp}} - D) = 0,
\end{aligned}
\end{equation}
\begin{equation}\label{4th-kkt-1}
\begin{aligned}
& \nabla_{ z, D_{\textup{cp}} } \mathcal{L}(z,D_{\textup{cp}},\Lambda) = \Big[\frac{\partial \mathcal{L}}{\partial z} ~ \frac{\partial \mathcal{L}}{\partial D_{\textup{cp}}} \Big]^\top = [0~0]^\top.
\end{aligned}
\end{equation}
\end{subequations}
\noindent where $\nabla$ is the gradient operator and $[\cdot]^\top$ is the transpose operator.

We first determine the optimal constellation size which will minimize \eqref{psi-dc-2} and then use it to get the optimal transmit power and compression ratio. It can be shown that $\fixwidetilde{\Psi}$ in \eqref{psi-dc-2} is convex in $z$ and there lies a global minima. From \eqref{4th-kkt-1} we have $\Big[\frac{\partial \mathcal{L}}{\partial z} ~ \frac{\partial \mathcal{L}}{\partial D_{\textup{cp}}} \Big]^\top = [0~0]^\top$. Taking partial derivative of \eqref{lag-main-1} with respect to $z$ and setting $\frac{\partial \mathcal{L}}{\partial z} = 0$ and after simplification we get
\begin{equation}\label{lag-eq-1-1}
 \frac { 3 \Omega T_\textup{s}   } {\mu  |h|^2}
  \Big( \textup{exp}\big(z/2\big) {-} 1 \Big) \Big( (z{-}1) \textup{exp}\big(z/2\big) {+} 1 \Big)  +  T_\textup{s} P_\textup{o}
  + \Lambda_{1} T_\textup{s}
  + \frac {  (\Lambda_{2} {-} \Lambda_{3}) z^2 \textup{exp}(z) } {D_{\textup{cp}} \ln (2)}
 =  0.
\end{equation}

Similarly, it can also be shown that $\fixwidetilde{\Psi}$ in \eqref{psi-dc-2} is convex in $D_{\textup{cp}}$ and there lies a global minima. Taking partial derivative of \eqref{lag-main-1} with respect to $D_{\textup{cp}}$ and setting $\frac{\partial \mathcal{L}}{\partial D_{\textup{cp}}} = 0$ and after simplification we get
\begin{equation}\label{lag-eq-2-1}
 \frac { 3 \Omega T_\textup{s}  } {\mu  |h|^2} \big(\textup{exp}\big(z/2\big) - 1\big)^2
 + T_\textup{s} P_\textup{o}
 + \Lambda_{1} \bigg( T_\textup{s} - \frac{z \tau \beta D^{\beta+1}  }{\ln (2) D_{\textup{cp}}^{\beta+1}}  \bigg)
 - \frac{\Lambda_{4} z}{\ln (2)}
 + \frac{\Lambda_{5} z}{\ln (2)}
 = \frac{z \tau \beta D^{\beta+1} P_\textup{cp}}{\ln (2)  D_{\textup{cp}}^{\beta+1}}.
\end{equation}

From complimentary slackness condition \eqref{3rd-kkt-1} we know either $\Lambda_{i}$ is zero or the associated constraint function is zero for any given $i$. First we consider one of the possible cases that is $\Lambda_{1}$, $\Lambda_{2}, \Lambda_{3}, \Lambda_{4}, \Lambda_{5}$ do not exist, i.e., unconstrained minimization. Accordingly, plugging in $\Lambda_{1} {=} 0, \Lambda_{2} {=} 0, \Lambda_{3} {=} 0, \Lambda_{4} {=} 0, \Lambda_{5} {=} 0$ in \eqref{lag-eq-1-1}, \eqref{lag-eq-2-1} yields following expressions, respectively,
\begin{equation}\label{lag-eq-3-1}
 \frac { 3 \Omega  } {\mu  |h|^2}
  \Big( \textup{exp}\big(z/2\big) {-} 1 \Big) \Big( (z{-}1) \textup{exp}\big(z/2\big) {+} 1 \Big)  + P_\textup{o}  =  0,
\end{equation}
\begin{equation}\label{lag-eq-4-1}
 - \frac {z \tau \beta D^{\beta+1} P_\textup{cp} } {T_\textup{s} D_{\textup{cp}}^{\beta+1} \ln (2)}
 + \frac { 3 \Omega} {\mu |h|^2} \Big(\textup{exp}\big(z/2\big) {-} 1\Big)^2
 +  P_\textup{o}  = 0.
\end{equation}

Solving \eqref{lag-eq-4-1} for $D_{\textup{cp}}$ yields
\begin{equation}\label{lag-eq-5-1}
D_{\textup{cp}} = D
                \Bigg(
                \frac {      z \tau \beta P_\textup{cp}
                      }
                      {
                        \frac { 3 \Omega T_\textup{s} \ln(2)  } {\mu |h|^2}
                        \big( \textup{exp}(z/2) {-} 1 \big)^2
                        {+} P_\textup{o} T_\textup{s} \ln(2)
                      }
                \Bigg)^{\frac{1}{\beta+1}}.
\end{equation}

Numerically solving \eqref{lag-eq-3-1} for $z$ yields its value $\tilde{z}$. Substituting this value of $z$ in \eqref{lag-eq-5-1} and solving for $D_{\textup{cp}}$ yields its value $\fixwidetilde{D}_\textup{cp}$. $\tilde{z}$ and $\fixwidetilde{D}_\textup{cp}$ provide a lower bound on optimization problem in \eqref{opt-prob-1-transformed} for given instantaneous CGI, $|h|^2$. It can be shown that $\tilde{z}$ and $\fixwidetilde{D}_\textup{cp}$ satisfy all the KKT conditions when the first constraint in \eqref{opt-prob-1-transformed} is slack, i.e.,
\begin{equation}\label{lag-eq-6-1}
\tau D^{\beta+1} \fixwidetilde{D}_\textup{cp}^{-\beta} - \tau D  +  \tilde{z}^{-1} \fixwidetilde{D}_\textup{cp} T_\textup{s} \ln(2)  - T < 0,
\end{equation}
and other constraints are also slack. Thus, the optimal lagrange multiplier $\Lambda_{1}$, $\Lambda_{2}, \Lambda_{3}, \Lambda_{4}, \Lambda_{5}$ are zero. Hence, the derived solution in \eqref{lag-eq-3-1} and \eqref{lag-eq-5-1} is the optimal solution for the optimization problem in \eqref{opt-prob-1-transformed} for given instantaneous CGI, $|h|^2$, when all constraints in \eqref{opt-prob-1-transformed} are slack.

Now consider another possible case when the first constraint in \eqref{opt-prob-1-transformed} is not slack, i.e., $\Lambda_{1}$ exits and $\Lambda_{2}, \Lambda_{3}, \Lambda_{4}, \Lambda_{5}$ do not exist. Thus, plugging in $\Lambda_{1} {\neq} 0, \Lambda_{2} {=} 0, \Lambda_{3} {=} 0, \Lambda_{4} {=} 0, \Lambda_{5} {=} 0$ in \eqref{lag-eq-1-1}, \eqref{lag-eq-2-1} and complimentary slackness conditions \eqref{3rd-kkt-1} yields following expressions, respectively,
\begin{equation}\label{lag-eq-7-1}
 \frac { 3 \Omega  } {\mu  |h|^2}
  \Big( \textup{exp}(z/2) {-} 1 \Big) \Big( (z{-}1) \textup{exp}\big(z/2\big) {+} 1 \Big)  + P_\textup{o}  + \Lambda_{1}  =  0,
\end{equation}
\begin{equation}\label{lag-eq-8-1}
 \Lambda_{1} \Big( 1  - \frac{z \tau \beta D^{\beta+1}  }{T_\textup{s} D_{\textup{cp}}^{\beta+1} \ln (2)}  \Big)
 - \frac {z \tau \beta D^{\beta+1}  P_\textup{cp} } {T_\textup{s} D_{\textup{cp}}^{\beta+1} \ln (2)}
 + \frac { 3 \Omega} {\mu |h|^2} \Big(\textup{exp}\big(z/2\big) {-} 1\Big)^2
 +  P_\textup{o}  = 0.
\end{equation}

Solving \eqref{lag-eq-7-1} for $\Lambda_{1}$ and substituting its value in \eqref{lag-eq-8-1} yields 
\begin{equation}\label{lag-eq-9-1}
\begin{aligned}
 \bigg(  \frac { 3 \Omega   } {\mu  |h|^2}
  \Big( \textup{exp}(z/2) {-} 1 \Big) \Big( (z{-}1) \textup{exp}(z/2) {+} 1 \Big) {+} P_\textup{o}  \bigg)
  \bigg( 1 {-} \frac{z \tau \beta D^{\beta+1} }{T_\textup{s} D_{\textup{cp}}^{\beta+1} \ln (2)}   \bigg)
 \\
 {+} \frac {z \tau \beta D^{\beta+1}  P_\textup{cp}} {T_\textup{s} D_{\textup{cp}}^{\beta+1} \ln (2)}
 {-} \frac { 3 \Omega} {\mu |h|^2} \big(\textup{exp}(z/2) {-} 1 \big)^2
 =  P_\textup{o}.
\end{aligned}
\end{equation}

Solving \eqref{lag-eq-9-1} for $D_{\textup{cp}}$ yields
\begin{equation}\label{lag-eq-10-1}
D_{\textup{cp}} = D \upsilon^{\frac{1}{\beta+1}},
\end{equation}
\noindent where
\begin{equation}
\upsilon =       \frac {
                             P_\textup{cp} {-} P_\textup{o}
                             {-} \frac { 3 \Omega  } {\mu |h|^2}  \big( \textup{exp}(z/2) {-} 1 \big) \big( (z{-}1) \textup{exp}(z/2) {+} 1 \big)
                      }
                      {
                        \frac { 3 \Omega  } {\tau \beta \mu |h|^2}
                        \big( \textup{exp}(z/2) - \textup{exp}(z) \big)
                      }.
\end{equation}
Substituting $D_{\textup{cp}}$ from \eqref{lag-eq-10-1} in \eqref{lag-eq-9-1} yields
\begin{equation}\label{lag-eq-11-1}
\frac{T}{D} + \tau - \tau \upsilon^{\frac{-\beta}{\beta+1}}  =  \frac{ T_{\textup{s}} \ln(2) } {z} \upsilon^{\frac{1}{\beta+1}}.
\end{equation}

Numerically solving \eqref{lag-eq-8-1} for $z$ yields its value $\hat{z}$. Substituting this value of $z$ in \eqref{lag-eq-7-1} and solving for $D_{\textup{cp}}$ yields its value $\fixwidehat{D}_\textup{cp}$. The value of $z$ can be obtained by numerically solving \eqref{lag-eq-8-1}. Substituting value of $z$ in \eqref{lag-eq-7-1} and solving for $D_{\textup{cp}}$ yields its value. It can be shown that $\hat{z}$ and $\fixwidehat{D}_\textup{cp}$ satisfy all the KKT conditions when all constraints in \eqref{opt-prob-1-transformed} are slack except the first constraint, thus the optimal lagrange multiplier $\Lambda_{1}$ is positive and $\Lambda_{2}, \Lambda_{3}, \Lambda_{4}, \Lambda_{5}$ are zero. Hence, the derived solution in \eqref{lag-eq-10-1} and \eqref{lag-eq-11-1} is the optimal solution for the optimization problem in \eqref{opt-prob-1-transformed}, when all constraints in \eqref{opt-prob-1-transformed} are slack except the first constraint. For all other cases, similar steps can be followed and it can be shown that these cases violate one or more constraints.

The problem in \eqref{opt-prob-1-transformed} is equivalent to \eqref{opt-prob-p-1}, thus the optimal values of $M$ and $D_{\textup{cp}}$ for both cases can be obtained by substituting $z=\ln(M)$ in \eqref{lag-eq-3-1}, \eqref{lag-eq-5-1} and \eqref{lag-eq-11-1}, \eqref{lag-eq-10-1}, respectively, which will minimize the objective function in \eqref{opt-prob-p-1}. Finally, by substituting the optimal value of $M$ in \eqref{min-Pt-propo-1} we can determine the optimal $P_{\textup{t}}$ which will minimize $\Psi$ for given instantaneous CGI, $|h|^2$.

\section{Proof of Proposition 2}\label{C}

Substituting BER constraint \eqref{ber} and SNR expression \eqref{snr} in second constraint in \eqref{opt-prob-p-3} yields
\begin{equation}\label{ber-3-1}
  \mathbb{P} \bigg\{ \omega_2 \; \textup{exp} \bigg( -  \frac { \omega_1 } {(M-1)} \frac {\kappa P_\textup{t} |h|^2}{\sigma^2 d^\alpha}  \bigg)
  \leqslant        \phi \bigg\} \geqslant \vartheta.
\end{equation}

Solving \eqref{ber-3-1} for fading power gain, $|h|^2$, yields
\begin{equation}\label{ber-3-3}
  \mathbb{P} \bigg\{ |h|^2 \geqslant (1-M) \frac{ \sigma^2 d^\alpha \ln ( \phi/\omega_2 ) } {   \omega_1 \kappa P_\textup{t} }  \bigg\} \geqslant \vartheta.
\end{equation}

The left hand side of \eqref{ber-3-3} represents the complimentary cumulative distribution function (ccdf) for $|h|^2$. For our considered Rayleigh fading channel, the fading power gain, $|h|^2$, is exponentially distributed. \eqref{ber-3-3} can be given as follows
\begin{equation}\label{ber-3-4}
  1 - \bigg[ 1 - \textup{exp} \bigg(  (M-1) \frac{ \sigma^2 d^\alpha \ln ( \phi/\omega_2 ) } {  \varsigma \omega_1 \kappa P_\textup{t} }  \bigg) \bigg] \geqslant \vartheta,
\end{equation}
\noindent where $\varsigma$ represents the scale parameter of the probability distribution. Solving \eqref{ber-3-4} for $P_{\textup{t}}$ yields
\begin{equation}\label{ber-3-5}
  P_\textup{t} \geqslant (M-1) \frac{ \sigma^2 d^\alpha \ln ( \phi/\omega_2 ) } { \varsigma \omega_1 \kappa \ln ( \vartheta) }.
\end{equation}

By substituting $T_{\textup{cp}}$, $T_{\textup{tx}}$, $P_\textup{tx}$, $r$ and $\varepsilon$ from \eqref{t-comp}, \eqref{t-comm}, \eqref{p-comm}, \eqref{r} and \eqref{mqam-par}, respectively, in \eqref{psi} yields $\Psi$ as a function of $M$, $D_{\textup{cp}}$ and $P_{\textup{t}}$, which can be expressed as follows
\begin{equation}\label{psi-dc-3-0}
  \Psi  =    \tau D^{\beta+1} D_{\textup{cp}}^{-\beta} P_{\textup{cp}} - \tau D P_{\textup{cp}}
                   +  \frac{ D_{\textup{cp}} T_\textup{s} \ln(2) }{  \ln(M) }
                       \bigg(  \frac{3 P_\textup{t} ({M}^{\frac{1}{2}}-1)}{\mu ({M}^{\frac{1}{2}}+1)}   + P_\textup{o} \bigg).
\end{equation}

For given values of $M$ and $D_{\textup{cp}}$, $\Psi$ is an increasing function of $P_{\textup{t}}$. Hence, the best choice of $P_{\textup{t}}$ to minimize $\Psi$ while satisfying the constraint in \eqref{ber-3-5} is the minimum value obtained by setting \eqref{ber-3-5} with equality. Thus, $P_{\textup{t}}$ can be expressed as a function of $M$, as given in \eqref{min-Pt-propo-3}.

\end{document}